%% file: lutp1818.tex
\documentclass[12pt,a4paper]{article}

\input{setup}

\begin{document}
\sloppy

\pagestyle{empty}

\begin{flushright}
LU TP 18-18\\
MCnet-18-20\\
August 2018\\
Revised November 2018
\end{flushright}

\vspace{\fill}
\begin{center}

{\Huge\bf The Space--Time Structure}\\[4mm]
{\Huge\bf of Hadronization in the Lund Model}\\[10mm]
{\Large Silvia Ferreres-Sol\'e\footnote{Now at NIKHEF, Science Park 105,
 NL-1098 XG Amsterdam, Netherlands} and Torbj\"orn Sj\"ostrand}\\[3mm]
{\it Theoretical Particle Physics,}\\[1mm]
{\it Department of Astronomy and Theoretical Physics,}\\[1mm]
{\it Lund University,}\\[1mm]
{\it S\"olvegatan 14A,}\\[1mm]
{\it SE-223 62 Lund, Sweden}
\end{center}

\vspace{\fill}

\begin{center}
\begin{minipage}{\abstwidth}
\begin{center} {\bf Abstract}
\end{center}
The assumption of linear confinement leads to a proportionality of the
energy--momentum and space--time pictures of fragmentation for a simple
$\q\qbar$ system in the Lund string model. The hadronization of more
complicated systems is more difficult to describe, and in the past only
the energy--momentum picture has been implemented. In this article
also the space--time picture is worked out, for open and closed
multiparton topologies, for junction systems, and for massive quarks.
Some first results are presented, for toy systems but in particular
for LHC events. The density of hadron production is quantified under
different conditions. The (not unexpected) conclusion is that this
density can become quite high, and thereby motivate the observed
collective behaviour in high-multiplicity $\p\p$ collisions. The new
framework, made available as part of the \textsc{Pythia} event
generator, offers a starting point for future model building in a
number of respects, such as hadronic rescattering.
\end{minipage}
\end{center}

\vspace{\fill}

\phantom{dummy}

\clearpage

\pagestyle{plain}
\setcounter{page}{1}

\section{Introduction}
\label{sec:introduction}

The Standard Model of particle physics is solidly established by now,
and has been very successful in describing all perturbatively
calculable observables for LHC $\p\p$ collisions, i.e.\ those
dominated by large momentum transfer scales \cite{Campbell:2017hsr}.
But at lower scales the perturbative approach breaks down, and
phenomenological models have to be developed.

One of the underlying assumptions for these models has been that the
nonperturbative hadronization process, wherein the perturbatively
produced partons turn into observable hadrons, is of a universal character.
Then relevant nonperturbative parameters can be determined e.g.\ from
LEP data, and afterwards be applied unmodified to LHC $\p\p$ collisions.
The hadronizing partonic state is quite different in the two processes,
however. Firstly, the composite nature of the incoming protons leads to
multiple semiperturbative parton--parton collisions, so-called MultiParton
Interactions (MPIs) \cite{Sjostrand:1987su,Sjostrand:2017cdm},
and also to beam remnants and initial-state QCD radiation. Secondly,
the high number of interacting partons leads to the possibility
of nontrivial and dynamically evolving colour topologies, collectively
referred to as Colour Reconnection (CR) phenomena. Both MPIs and CR need
to be modelled, and involve further new parameters. (CR has been observed
in the cleaner $\e^+ \e^- \to \W^+ \W^-$ process by the LEP collaborations
\cite{Schael:2013ita}, but that information is not easily transposed
to the $\p\p$ context.)

The most successful approach to providing a combined description of all
relevant phenomena, at all scales, is that of event generators.
Here Monte Carlo methods are used to emulate the quantum mechanical
event-by-event fluctuations at the many stages of the evolution of an event
\cite{Buckley:2011ms}. For $\p\p$ physics the three most commonly used
generators are \textsc{Pythia} \cite{Sjostrand:2006za,Sjostrand:2014zea},
Herwig \cite{Bahr:2008pv,Bellm:2015jjp} and \textsc{Sherpa}
\cite{Gleisberg:2008ta}. Fragmentation here proceeds either via strings
\cite{Andersson:1983ia}, for the former, or via clusters
\cite{Webber:1983if}, for the latter two. A note on terminology:
``fragmentation'' and ``hadronization'' can be used almost interchangeably,
but the former is more specific to the breakup of a partonic system
into a set of primary hadrons, whereas the latter is more generic
and can also include e.g.\ decays of short-lived resonances.

In spite of an overall reasonable description, glaring discrepancies
between data and models have been found in some cases. Most interesting is
that high-multiplicity LHC $\p\p$ events show a behaviour that
resembles the one normally associated with heavy-ion collisions and the
formation of a Quark-Gluon Plasma (QGP). In particular, ALICE has shown
that the fraction of strange baryons increases with multiplicity,
the more steeply the more strange quarks the baryon contains, while the
proton rate is not enhanced \cite{ALICE:2017jyt}. Long-range azimuthal
``ridge'' correlations have also been observed by both CMS
\cite{Khachatryan:2010gv,Khachatryan:2015lva} and ATLAS
\cite{Aad:2015gqa}, as well as other signals of collective flow
\cite{Ortiz:2013yxa,Abelev:2014qqa,Khachatryan:2016txc}.

This is unlike conventional expectations, that QGP formation requires
volumes and timescales larger than the one that can be obtained in $\p\p$
collisions \cite{BraunMunzinger:2007zz,Busza:2018rrf,Nagle:2018nvi}.
Nevertheless core--corona models have been developed, like the one
implemented in EPOS \cite{Werner:2014xoa}, where a central high-density
region can turn into a QGP, while the rest of the system remains as normal
individual strings. Other mechanisms that have been proposed include
rope formation \cite{Bierlich:2014xba} and shoving \cite{Bierlich:2016vgw},
or an environment-dependent string tension \cite{Fischer:2016zzs}.
Common  for all of them is that they introduce a space--time picture
of the collision process.

In the traditional Lund string model \cite{Andersson:1983ia} the linear
confinement potential leads to a linear relationship between the
energy--momentum and space--time pictures of a simple $\q\qbar$
fragmenting system. Many of the above models are based on the
approximation of a number of such simple strings, parallel along the
$\p\p$ collision axis but displaced in the transverse plane by the
collision/MPI geometry.

For a generic multiparton system, like $\q\g_1\g_2\ldots\g_n\qbar$,
only an energy--momentum picture has been available until now
\cite{Sjostrand:1984ic}. The purpose of this article is to overcome
that limitation, and provide a full space--time picture of the
hadronization process, as part of the \textsc{Pythia} event generator.%
\footnote{After posting the preprint, we learned of another related
implementation \cite{Gallmeister:2005ad}. It does not address all the 
issues considered here, however, and therefore is insufficient e.g.\
for LHC studies.}
This will offer a natural starting point for more detailed future
studies of a number of collective effects. The models mentioned above
deal with the space--time structure before (like core--corona or shove)
or during (like ropes or QGP) fragmentation. To this we would also add
a possibility for studies of what happens after the first stages of the
hadronization, when hadronic rescattering and decays can occur in parallel.
In addition to the already mentioned observables, Bose--Einstein correlations
could also be used to characterize final states.

A warning is that we are applying semiclassical models to describe
the quantum world. Formally the Heisenberg uncertainty relations
impose limits on how much simultaneous energy--momentum and space--time
information one can have on an individual hadron. Our approach should
still make sense when averaged over many hadrons in many events, as will
always be the case.

The plan of the article is as follows.
Section~\ref{sec:lundstring} gives a brief summary of relevant earlier
work, on the ``complete''  description of the simple $\q\qbar$ system
\cite{Andersson:1983ia}, and on the energy--momentum picture of an
arbitrary partonic system \cite{Sjostrand:1984ic}.
Section~\ref{sec:spacetime} then introduces the new framework that
provides a space--time picture also in a general configuration. Several
special cases need to be addressed, and technical complications have to
be sorted out, with some details relegated to two appendices.
Section~\ref{sec:hadrondensity}  contains some first studies, partly
for toy systems but mainly for LHC events. This is  without any of the
collective effects that may be added later, but still provides an
interesting overview of the overall space--time evolution of hadronization
at the LHC. Finally, Section~\ref{sec:summary} concludes with a summary
and outlook.

Natural units are assumed throughout the article, i.e.\ $c = \hbar = 1$.
By default energy, momentum and mass is given in GeV, and space and
time in fm.

\section{The Lund String Model}
\label{sec:lundstring}

\begin{figure}[tp]
\centering
\includegraphics[scale=0.45]{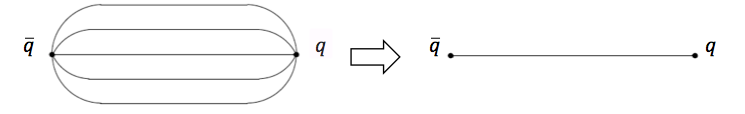}
\caption{A simplified colour-field topology in a $\q\qbar$ system and
its further simplified string representation.}
\label{fig:tubelike}
\end{figure}

\subsection{The linear force field in QCD}
\label{sec:linearforce}

Confinement is one of the most fundamental properties of QCD.
It can be viewed as a consequence of an approximately linear term
in the QCD potential
\clearpage
\begin{equation}
V_{\mathrm{QCD}}(r) \approx -\frac{4}{3} \, \frac{\as}{r} + \kappa \, r
\label{eq:QCDpotential}
\end{equation}
between a quark and an antiquark in an overall colour singlet state,
where $r$ is the distance between them and $\as$ is the strong coupling
constant. The presence of a linear term was first inferred from hadron
spectroscopy (Regge trajectories), from which a $\kappa \approx 1$~GeV/fm
can be extracted, and has later been confirmed by lattice QCD
calculations.

The linear term dominates at large distances, and in the Lund
string model only this term is used to describe the breakup of a
high-mass $\q\qbar$ system into several smaller-mass ones. Then the
full colour field can be approximated by a one-dimensional string
stretched straight between the $\q$ and $\qbar$, Fig.~\ref{fig:tubelike}.
This string can be viewed as parametrizing the center of a cylindrical
region of uniform width along its full length, such that the longitudinal
and transverse degrees of freedom almost completely decouple.

\subsection{The two-parton system}
\label{sec:twoparton}

The Lund model is easiest to understand in the context of a simple
quark--antiquark pair created at the origin (e.g.\ by $\e^+ \e^-$
annihilation) and moving out along the $\pm z$ axis. Neglecting
the transverse degrees of freedom, the Hamiltonian can then be written
as \cite{Andersson:1983ia}
\begin{equation}
H = E_{\q} +E_{\qbar} + \kappa |z_{\q}-z_{\qbar}| ~.
\label{eq:qqbarhamiltonian}
\end{equation}
Here $|z_{\q}-z_{\qbar}|$ is the distance between $\q$ and $\qbar$,
and $E_{\q}$ and $E_{\qbar}$ are the energies of the $\q$ and $\qbar$.
With both assumed massless, it also holds that
$E_{\q/\qbar}= |\bp_{\q/\qbar}|=|p_{z,\q/\qbar}|$.

From the Hamiltonian, the equation of motion gives rise to a linear
relation between the space--time and the energy--momentum pictures
\begin{equation}
\left| \frac{\d p_{z,\q/\qbar}}{\d t} \right| =
\left| \frac{\d p_{z,\q/\qbar}}{\d z} \right| =
\left| \frac{\d E_{\q/\qbar}}{\d t} \right| =
\left| \frac{\d E_{\q/\qbar}}{\d z} \right| = \kappa ~.
\label{eq:xplinearity}
\end{equation}
The signs of the derivatives depend both on the direction of motion 
of the parton and on the direction the string pulls it in. When
the parton moves out along the $+z$ axis, e.g.,  the string pulls the 
parton in the $-z$ direction, and all signs are negative.

\subsubsection{Simple string motion}
\label{sec:stringmotion}

\begin{figure}[tp]
\centering
\subfloat[$\q\qbar$ system in the CM frame.]{
    \includegraphics[scale=0.48]{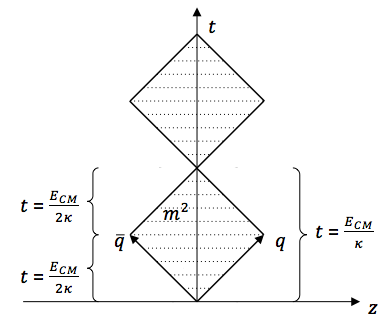}}
\hspace{2mm}
\subfloat[$\q\qbar$ system in a boosted frame with respect to
the CM frame.]{
    \includegraphics[scale=0.5]{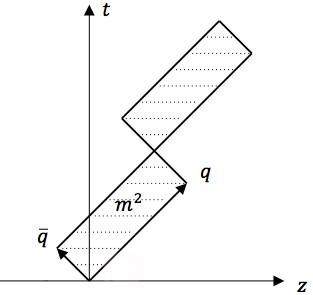}}
\caption{The motion of a $\q\qbar$ system, with massless $\q$ and $\qbar$.}
\label{fig:qqbarsystem}
\end{figure}

In the absence of string breaks, the motion of the simple $\q\qbar$
system in its rest frame can be described as a ``yo-yo'' motion, where
a string is alternatingly ``reeled out'' and ``reeled in'',
Fig.~\ref{fig:qqbarsystem}a. In the first quarter of a period the
$\q$ and $\qbar$ are moving apart from each other with the speed of light,
$z = \pm t$, such that the string length $l_{\mathrm{string}} = 2t$.
Therefore, the four-momenta of the $\q$, $\qbar$ and the string evolve
with time as
\begin{equation}
\begin{gathered}
p_{\q/\qbar}(t) = \left( \frac{\ECM}{2}-\kappa t \right)
\, (1; 0, 0, \pm 1) ~,\\
p_{\mathrm{string}}(t) = (2\kappa t; 0, 0, 0) ~,
\end{gathered}
\label{eq:momentumatt}
\end{equation}
where $\ECM$ is the center-of-mass energy of the full system. At time
$t = \ECM/2\kappa$ all the energy is carried by the string, whose string
tension then forces the $\q$ and $\qbar$ to turn around. In the second
quarter of the period the string length decreases like
$l_{\mathrm{string}} = 2(\ECM/\kappa - t)$, and energy and momentum is
transferred back to the $\q / \qbar$. At $t = \ECM/\kappa$ the string
has vanished and the $\q / \qbar$ are back at the origin, but now
moving in the $\mp z$ direction. The second half of the full period
therefore becomes a repeat of the first half, only with the role
of $\q$ and $\qbar$ interchanged. Normally string breaks are assumed to
occur so rapidly that only the first quarter of the first period needs to
be considered.

The kinematics of the yo-yo motion can conveniently be rewritten in
terms of light-cone coordinates, both in energy--momentum,
$\tdp^{\pm} = E \pm p_{z}$, and in space--time, $\tilde{z}^{\pm} = t \pm z$.
For instance, for the quark in the first quarter period,
$\tilde{z}_{\q}^- = \tdp_{\q}^- =  0$, $\tilde{z}_{\q}^+ = 2t$,
$\tdp_{\q}^+ = \ECM - 2 \kappa t = \ECM - \kappa \tilde{z}_{\q}^+$,
such that $\d\tdp_{\q}^+ / \d\tilde{z}_{\q}^+ = -\kappa$.
$\tdp^{\pm}$ also obey the relation
\begin{equation}
\tdp^+ \tdp^- = (E+p_{z})  (E-p_{z}) = m^2 + p_x^2 + p_y^2
= m^2 + \pT^2 = \mT^2 ~,
\label{eq:massofhadron}
\end{equation}
which reduces to $\tdp^+ \tdp^- = m^2$ when $p_x = p_y = 0$.

The simplest yo-yo system can be generalized as illustrated in
Fig.~\ref{fig:qqbarsystem}b, where the quark and the antiquark have
different initial energies, $E_{\q} \neq E_{\qbar}$. Equivalently,
this system can be viewed as a boosted copy of the rest-frame setup in
Fig.~\ref{fig:qqbarsystem}a. The energy--momentum and space--time
coordinates suffer simultaneous transformations under a longitudinal
boost, and eq.~(\ref{eq:xplinearity}) holds also after the boost.
The transformation is especially easily formulated in light-cone
coordinates, where $\tdp'^{\pm} = k^{\pm 1} \tdp^{\pm}$ with
$k = \sqrt{(1 + \beta) / (1 - \beta)}$ for a boost with velocity
$\beta$, and similarly for $\tilde{z}^{\pm}$.

Note that a string piece with $E = \kappa l$ but $p_z = 0$ in the
original rest frame will obtain a $p_z \neq 0$ after a boost to the
frame with $E_{\q} \neq E_{\qbar}$. This is in seeming contradiction
with a description set up in a rest frame where $E_{\q} \neq E_{\qbar}$
from the onset, where one would again expect $p_z = 0$. The solution is
that a string piece is an extended object, so that the two ends of it,
if originally simultaneous, will no longer be it after the boost.
Only a string piece at constant time in the new frame will obey
$E = \kappa l$ and $p_z = 0$ there.

\subsubsection{String breaking and hadron formation}
\label{sec:stringbreaking}

As described in the previous section, the potential energy stored in
the string increases with the separation between an original $\q_0$
and $\qbar_0$. This makes the creation of a new $\q_1\qbar_1$ pair
in the string energetically favourable, if the invariant mass of the
system is big enough. It is here assumed that colours are matched so
that the original colour-singlet $\q_0\qbar_0$ string breaks into
two pieces, $\q_0\qbar_1$ and $\q_1\qbar_0$, that separately are colour
singlets. By local flavour conservation the
$\q_1$ and $\qbar_1$ have to be created in the same vertex. They are
created with vanishing energy, and are then pulled apart by the string
tension. Naively, the probability for the string to break increases
with time, because the string length increases. On the other hand,
a break can inhibit later breaks, since each break fragments the string
into two smaller systems, leaving an in-between region without a string.
If on-mass-shell criteria for hadrons are ignored, as in the
Artru-Mennessier model \cite{Artru:1974hr}, a naive constant breakup
probability per unit string area then is modified by an
exponential-decay factor.

\begin{figure}[tp]
\centering
\subfloat[$\hx^{\pm}$  and $x^{\pm}$ fractions]{
    \includegraphics[scale=0.35]{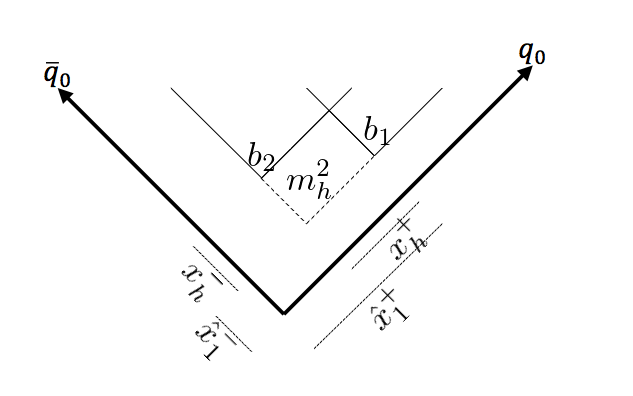}}
\hspace{2mm}
\subfloat[$z^{\pm}$  fractions]{
    \includegraphics[scale=0.35]{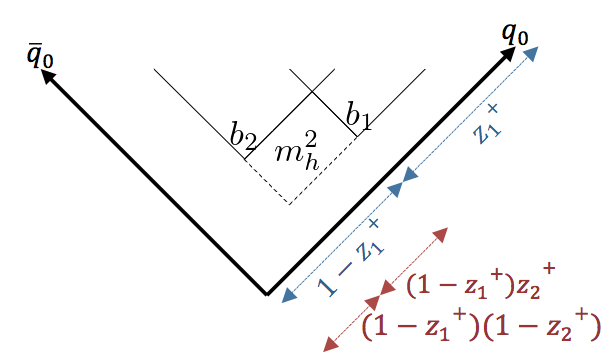}}
\caption{Simple $\q\qbar$ system, where $\q$ and $\qbar$ are massless,
with two breaks, $b_1$ and $b_2$. The light-cone coordinates
are normalized to unity.}
\label{fig:xzfractions}
\end{figure}

In general, several string breaks can occur between the $\q_0$
and $\qbar_0$. Consider two adjacent ones, $b_1$ at $(t_1, z_1)$ and
$b_2$ at $(t_2, z_2)$, as depicted in Fig.~\ref{fig:xzfractions}.
The $\q_1$ from the $b_1$ vertex combines with the $\qbar_2$ from
the $b_2$ vertex, forming a hadron $\q_1\qbar_2$ with mass $m_h$.
Since $\q_1$ and $\qbar_2$ are created with no energy--momentum,
the four-momentum of the hadron entirely comes from the intervening
string piece, which can be read off like
\cite{Andersson:1983ia}
\begin{equation}
E_h = \kappa (z_1-z_2) ~,~~~ p_{zh} = \kappa  (t_1-t_2) ~,~~~
\tdp_h^{\pm} = \kappa \left| \tilde{z}_1^{\pm} - \tilde{z}_2^{\pm} \right| ~.
\label{eq:ephadroninst}
\end{equation}

Next consider the quantities $\hx^{\pm}$ and $x^{\pm}$, illustrated in
Fig.~\ref{fig:xzfractions}a. The former represent the light-cone
coordinates of breakup vertices, scaled by the corresponding coordinates
of the $\q_0$ and $\qbar_0$ turning points, so as to be restricted to a
physical region $0 \leq x^{\pm} \leq 1$. The latter represents the
light-cone separation between two (adjacent) breaks, correspondingly
scaled. For the $\q_1\qbar_2$ hadron this translates to
$x_h^+ = \hx_1^+ - \hx_2^+$, $x_h^- = \hx_2^- - \hx_1^-$. Defining the
two four-vectors $p^+ = p_{\q_0}(t=0) = E_{\q_0}(1; 0, 0, 1)$ and
$p^- = p_{\qbar_0}(t=0) = E_{\qbar_0}(1; 0, 0, -1)$, and using the
proportionality between space--time and energy--momentum, the hadron
four-momentum then becomes
\begin{equation}
p_h =  x_h^+ \, p^+ + x_h^- \, p^- ~.
\label{eq:fourpgeneral}
\end{equation}
Although eq.~(\ref{eq:fourpgeneral}) has been derived for a system in
which $\q$ and $\qbar$ are moving in opposite directions, it is valid
in all frames, which makes the $\hx^{\pm}$ coordinates and $x^{\pm}$
fractions most useful. Since $\ECM^2 = (p^+ + p^-)^2 = 2p^+p^-$, the
hadron mass obeys
\begin{equation} m_h^2 = p^{2}_h= \left( x_h^+ \, p^+ + x_h^- \, p^- \right)^2
= x_h^+x_h^- \, 2 p^+ p^- = x_h^+x_h^- \, \ECM^2~.
\label{eq:massinxfractions}
\end{equation}
Do note the factor of 2 for the $p^{\pm}$ vectors in $\ECM^2 = 2p^+p^-$,
as opposed to the relation $\ECM^2 = \tdp^+ \tdp^-$ for the two scalar
quantities $\tdp^{\pm}$, and correspondingly for the hadronic subsystems.

Each breakup vertex is characterized by its invariant time $\tau$.
A convenient corresponding energy--momentum quantity is
\begin{equation}
\Gamma =  (\kappa \tau) ^2 = \kappa^2 (t^2 - x^2 -y^2 -z^2) .
\label{eq:gamma}
\end{equation}
which geometrically corresponds to the string area in the backwards
light cone of the vertex. Using the notation of Fig.~\ref{fig:xzfractions}a
it can also be expressed as
\begin{equation}
\Gamma = (\hx^+ \, p^+ + \hx^-\, p^-) ^2 = \hx^+ \, \hx^- \, \ECM^2.
\label{eq:gammaep}
\end{equation}

\subsubsection{Selection of breakup vertices}
\label{sec:howpointsobtained}

The breakup vertices are causally disconnected. That is, $b_1$ and
$b_2$ in Fig.~\ref{fig:xzfractions} have a spacelike separation.
Which happens first then depends on the Lorentz frame in which the
event is studied. It is therefore possible to describe the fragmentation
process starting from the hadron closest to the $\q_0$ end and then
moving towards the $\qbar_0$ one, or the other way around. Assuming
e.g.\ that $b_1$ has already been selected, so that $\hx_1^{\pm}$
are fixed, then the selection of the two $x_h^{\pm}$ values of the
hadron defines the location of $b_2$. But, assuming that the hadron
and its mass are already specified, the mass constraint in
eq.~(\ref{eq:massinxfractions}) reduces it to one degree of freedom.
For the fragmentation from the $\q_0$ side this is conveniently
chosen to be the $x_h^+$ values. More specifically, $z^+$ fractions
are introduced, as illustrated in Fig.~\ref{fig:xzfractions}b,
as the hadron fraction of whatever system light-cone
momentum that still remains after the production of previous hadrons.
That is, the first hadron $\q_0 \qbar_1$ acquires a fraction
$z_1^+ = x^+_1$ of the total $\tdp^+$ of the system, while the
remnant-system is left with a $\tdp^+$ fraction of $1 - z_1^+ = 1 - x_1^+$.
The second hadron $\q_1\qbar_2$ takes a fraction $z_2^+$ from the
leftover $\tdp^+$, i.e.\ $x_2^+ = z_2^+ \, (1-z_1^+)$, leaving a new
remainder fraction $(1-z_1^+)(1-z_2^+)$. Since the fragmentation
process is iterative, the $x^{\pm}$ fractions related to hadron $i$
can be written as
\begin{equation}
\begin{gathered}
x_i^+ = z_i^+ \, \prod_{j=1}^{i-1} (1-z_{j}^+) ~, \\
x_i^-= \frac{m_i^2}{x_i^+ \, \ECM^2} ~,
\end{gathered}
\label{eq:relationzx}
\end{equation}
where the relation for $x_i^-$ is given by
eq.~(\ref{eq:massinxfractions}).

Alternatively the fragmentation could have been described from the
$\qbar_0$ end in terms of negative light-cone fractions $z^-$ and $x^-$.
Since the breakup points are causally disconnected, the two procedures
should result in the same average particle distribution. This
requirement, ``left--right symmetry'', gives a probability distribution
\cite{Andersson:1983ia,Andersson:1983jt}
\begin{equation}
f(z) \propto \frac{(1-z)^a}{z} \, \exp\left( -b\frac{m_h^2}{z} \right) ~,
\label{eq:probabilityofz}
\end{equation}
for the $z$ value of each new hadron, where $z = z^+$ ($z = z^-$) for
fragmentation from the $\q_0$ ($\qbar_0$) end. The $a$ and $b$ are
parameters that should be tuned to reproduce the experimental data.
Hence, $f(z)$ determines how the individual vertices correlate in order
to create a hadron of mass $m_h$ by taking a fraction $z$ of the
energy--momentum left in the system. Note that the form of $f(z)$
does not depend on previous steps taken, which leads to a flat rapidity
plateau of the inclusive hadron production.

The $\Gamma$ values of breakup vertices can be obtained recursively
by simple geometrical considerations,
\begin{equation}
\Gamma_i = (1-z_i) \left( \Gamma_{i-1} + \frac{m_i^2}{z_i} \right) ~,
\label{eq:newoldgammas}
\end{equation}
where $\Gamma_i$ and $\Gamma_{i-1}$ are the scaled squared invariant times
of the $i$ and $i-1$ breakups, respectively. The $\q_0$ and $\qbar_0$ turning points
define $\Gamma_0 = 0$. The inclusive $\Gamma$ distribution, after some
steps away from the endpoint(s), is
\begin{equation}
P(\Gamma) \propto \Gamma^a \, \exp{(-b\Gamma)}
\label{eq:probabilityofgamma}
\end{equation}
with the same $a$ and $b$ as in eq.~(\ref{eq:probabilityofz}).

\begin{figure}[tp]
\centering
\captionsetup{justification=centering}
\subfloat[First breakup from the $\q_0$ endpoint]{
    \includegraphics[scale=0.28]{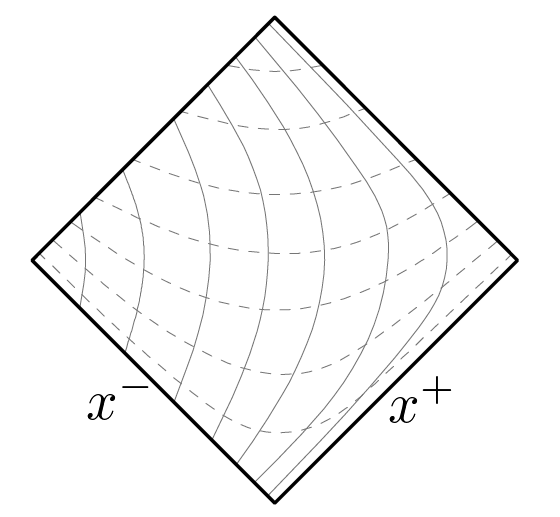}}
\hspace{2mm}
\subfloat[Some breakup inside the system]{
    \includegraphics[scale=0.28]{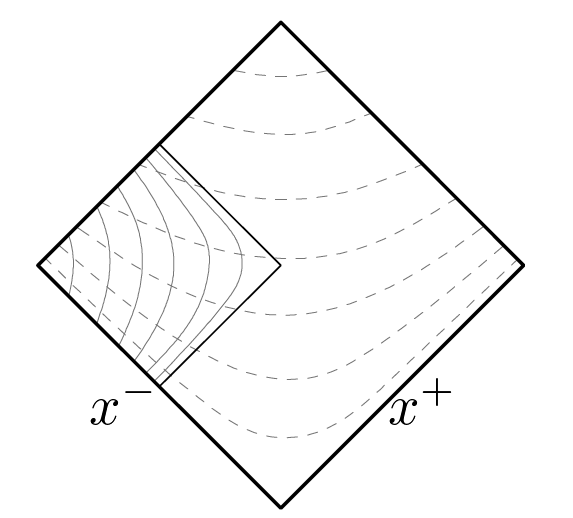}}
\caption{Hyperbolae of constant $\Gamma$ and $m_h^2$ represented
by dashed and full lines, respectively.}
\label{fig:breakuphyperbolae}
\end{figure}

The breaks of the string can be determined from eq.~(\ref{eq:relationzx})
by iteratively picking $z_i$ values according to
eq.~(\ref{eq:probabilityofz}) for the hadrons with masses $m_i$ .
This works well for the simple $\q_0 \qbar_0$ system, but
eq.~(\ref{eq:relationzx}) will not hold in systems with more than
two partons. To this end an alternative procedure can be introduced
\cite{Sjostrand:1984ic} via $\Gamma$ recursion. Here a $z = z_i$ is
still selected by eq.~(\ref{eq:probabilityofz}), and converted to a
$\Gamma_i$ by eq.~(\ref{eq:newoldgammas}). As illustrated in
Fig.~\ref{fig:breakuphyperbolae}, each fixed $\Gamma$ value corresponds
to a hyperbola with the origin as its center. Correspondingly each fixed
$m_i$ corresponds to a hyperbola with the $i-1$ vertex as its center.
Therefore a given $(m_i, \Gamma_i)$ pair corresponds to the unique crossing
of two hyperbolae at the location of the next vertex.

\subsubsection{The tunneling process}

Up to this point, we have assumed that the $\q_i\qbar_i$ pairs
generated from string breakings are massless and have no transverse
momenta. Both $\q_i$ and $\qbar_i$ are then created as real
particles at a common space--time location, with vanishing
energy--momentum. If the pair is massive or carries transverse
momentum, the $\q_i$ and $\qbar_i$ still have to be created in the
same space--time location, but as virtual particles. Each now has
to tunnel out a distance $l = \mT / \kappa$ to acquire enough energy
from the string to correspond to its transverse mass $\mT$.
This tunneling results in a Gaussian suppression factor
\begin{equation}
\exp\left(- \frac{\pi \mT^2}{\kappa} \right) =
\exp\left(-\frac{\pi m^2}{\kappa} \right)
\exp\left(-\frac{\pi \pT^2}{\kappa} \right) ~.
\end{equation}
A consequence of this mechanism is the suppression of heavy
quark production in string breaks, approximately like
$\u\ubar : \d\dbar : \s\sbar : \c\cbar \approx 1 : 1 : 0.3 : 10^{-11}$
\cite{Andersson:1983ia}. It is therefore assumed that
$\c$ and $\b$ production only occurs by perturbative processes.

The combination of a $\q_{i-1}$ and a $\qbar_i$ gives the flavour of a
meson but does not fully specify it. The quark spins can combine
e.g.\ to produce a pseudoscalar or a vector meson, and flavour-diagonal
mesons mix, and so on. All of these aspects are relevant for the model
as a whole, but for the considerations in this article we only need
to know the masses of the produced mesons. Similarly for baryon production,
where the production mechanisms are less well understood, whether
``diquark'' or ``popcorn'' \cite{Andersson:1984af,Eden:1996xi}.
In the latter approach actually three different production vertices
are involved, one for each of the quarks, but also here an effective
description in terms of two, as for mesons, is meaningful. A diquark
is taken to be a colour antitriplet, just like an antiquark, and we
thus use the notation $\qbar$ as shorthand for either of them.

Since the string itself has no transverse motion, it is assumed that
the transverse momentum is locally compensated inside each $\q_i\qbar_i$
pair. The transverse momentum of a hadron $\q_{i-1}\qbar_i$ is then
given by the vector sum of its constituent transverse momenta. The
hadron masses in section~\ref{sec:stringbreaking} have to be
replaced by the corresponding transverse masses.

\subsubsection{Massive quarks}
\label{sec:massivequarks}

Although massive quarks are not created from string breaking, they can
be generated in the hard process and form a system that might fragment
further. In this section, the yo-yo model is extended to account for
massive quarks as the endpoints of the system. Since the massive $\q$
and $\qbar$ do not travel at the speed of light, their motion is
described by hyperbolae instead of straight lines.

\begin{figure}[tp]
\centering
\subfloat[Motion of the $\c\cbar$ system.]{
    \includegraphics[scale=0.32]{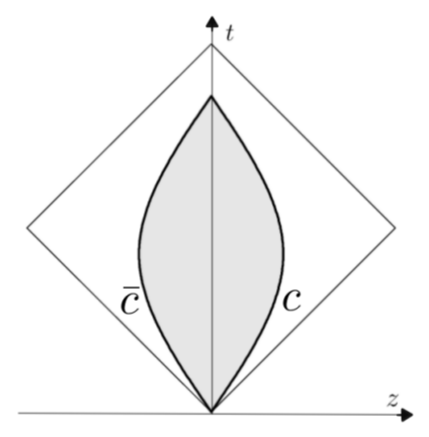}}
\hspace{10mm}
\subfloat[Location of the $\c\cbar$ system in the equivalent
massless $\q\qbar$ system.]{
    \includegraphics[scale=0.27]{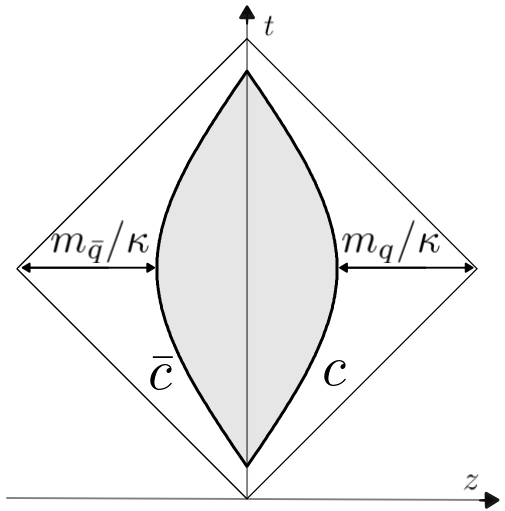}}
\caption{The $c\overline{c}$ system and the equivalent system formed
by massless $\q$ and $\qbar$.}
\label{fig:massivequarkorigin}
\end{figure}

To study the motion of the massive yo-yo system, consider a $\c\cbar$
system in the CM frame, in which $\c$ and $\cbar$ are moving along the
$z$ axis in opposite directions. The massive yo-yo system is depicted in
Fig.~\ref{fig:massivequarkorigin}a, along with the massless case for
comparison. At time $t = 0$
\begin{equation}
E_{\c}(0) = E_0 = \frac{\ECM}{2} ~,~~~
p_{z,\c}(0) = p_0 = \sqrt{E_0^2 - m_{\c}^2} ~.
\end{equation}
The proper relativistic definition of force, $\d p_{z} / \d t = \pm \kappa$,
then gives
\begin{equation}
\begin{gathered}
p_{z,\c}(t) = p_0 - \kappa t ~,\\
E_{\c}(t) = \sqrt{p_{z,\c}^2(t) + m_c^2} ~,\\
z_{\c}(t)  = \frac{E_0 - E_{\c}(t)}{\kappa} ~,
\end{gathered}
\end{equation}
with the motion of the $\cbar$ its mirror image. Notice that the
oscillation time is reduced by a factor $p_0 / E_0$ relative to the
massless system with the same $E_0$.

Although the motion properties of the massless and massive cases hold
in every longitudinal boosted frame, the effects of boosts are
simpler to address for massless quarks. A useful trick is to replace
the effect of the quark mass by an extra string piece of length
$(E_{\c}(t) - p_{z,\c}(t)) / \kappa$ at each endpoint.
Its length is $m_{\q}/\kappa$ at the turning point,
see Fig.~\ref{fig:massivequarkorigin}b, where the massive motion is
illustrated by the hyperbolae, whose asymptotes are the straight lines
of the massless case. Thereby time $t = 0$ is also offset to account
for the reduced oscillation time. The extra string piece is purely
fictitious and does not break during the fragmentation process.
The physical region, between the hyperbolae, is highlighted in grey
in Fig.~\ref{fig:massivequarkorigin}b. Given that the hadron created
from the endpoint is always heavier than the endpoint quark,
all the hadrons are automatically created inside the physical region.

The four-momenta of the massless reference four-vectors have to be
linear combinations of the massive quark four-momenta for Lorentz
covariance reasons. If $p_{\q}$ and $p_{\qbar}$ are the four-momenta of
the massive quarks, while $p_{0\q}$ and $p_{0\qbar}$ are the massless
four-momenta, the relation between them becomes
\begin{equation}
\begin{gathered}
p_{0\q} =  (1+k_1) p_{\q} -k_2p_{\qbar} ~,\\
p_{0\qbar} =  (1+k_2) p_{\qbar} - k_1p_{\q} ~,
\end{gathered}
\label{eq:masslessrefvec}
\end{equation}
where the $k_1$ and $k_2$ values are fixed by
$p_{0\q}^2 = p_{0\qbar}^2 = 0$ \cite{Sjostrand:1984ic}.

\subsection{Multiparton systems}
\label{sec:multipartonsystem}

Next, more complicated string topologies need to be considered.
An example is the $\Z^{0}$ decay into a pair of massless quarks,
either of which can emit a gluon:
\begin{equation*}
\Z^{0} \to \q\qbar \to \q\g\qbar ~.
\end{equation*}
Both such radiation and the hadronization can occur over widely
varying time scales in high-energy events, but in a local context
the radiation takes place at time scales shorter than those of the
hadronization itself. As a reasonable first approximation all three
partons can thus be assumed created at the origin.

In the Lund model the colour flow is based on the limit of infinitely
many colours \cite{tHooft:1973alw}. Then there is one string piece from
the $\q$ to the $\g$ and another from the $\g$ to the $\qbar$, and
the two do not interfere. The gluon thus can be viewed as a kink
on a single string stretched from the $\q$ to the $\qbar$.

\begin{figure}[tp]
\centering
\includegraphics[scale=0.6]{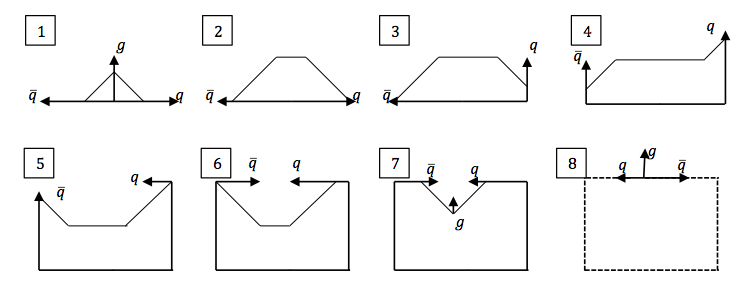}
\caption{Time evolution of the $\q\g\qbar$ system formed by massless
partons in a frame where the gluon moves in the $+x$ direction, while
the $\q$ and $\qbar$ move in opposite directions along the $z$ axis.}
\label{fig:qqgevolution}
\end{figure}

The motion of the $\q\g\qbar$ string system can conveniently be
studied in a Lorentz frame where the $\q$ moves in the $+z$ direction,
$\g$ in the $+x$ direction and $\qbar$ in the $-z$ direction. Two
string pieces are present initially, as illustrated in view 1 of
Fig.~\ref{fig:qqgevolution}. Each string piece defines a separate
string region, which behaves similarly to the string piece of a
$\q\qbar$ system, except that it is now transversely boosted.
The region formed by the $\q\g$ string  evolves with time as
\begin{equation}
\begin{gathered}
p_{\q}(t) = (E_{\q}(0) - \kappa t)  (1; 0, 0, 1) ~, \\
p_{g}(t) = (E_{\g}(0) - 2\kappa t)  (1; 1, 0, 0) ~,\\
p_{\mathrm{string}}(t) = \kappa t (2;  1,  0,  1) ~,
\label{eq:qgqatt}
\end{gathered}
\end{equation}
and correspondingly for $\qbar\g$, but with $z \to -z$.
Note the factor of 2 in the gluon four-momentum, which comes
from the loss of energy--momentum to both string pieces attached to it.
Unlike the simple $\q\qbar$ system, the two string pieces are not at
rest, but move in the transverse direction: the $\q\g$ string piece
has a velocity vector $v_x = v_z = 1/2$, while for the $\g\qbar$ piece
$v_x = -v_z = 1/2$. The energy per unit string length is higher than
for a string at transverse rest, but the lower string length drawn
out per unit time exactly compensates, such that the force acting on
the endpoints is of the same magnitude \cite{Sjostrand:1984ic}.

After time $t = E_{\g}(0)/2 \kappa$ the gluon has lost all its energy and
a new string piece is created by the inflowing momentum from the $\q$ and
$\qbar$, and is hence denoted as the $\q\qbar$ region, see view 2.
Later the $\q$ has also lost all its energy and starts to move in
the $+x$ axis as it absorbs $\g$ four-momentum. The $\q$ eventually gains
and re-emits half of the gluon energy, views 3 and 4. Subsequently it
absorbs original $\qbar$ four-momentum and moves in the $-z$ direction,
views 5 and 6. A similar process occurs for $\qbar$. As shown in view 7,
the gluon will eventually reappear, and in view 8 the sequence starts to
repeat, only with the momenta of $\q$ and $\qbar$ swapped.

\begin{figure}[tp]
\centering
\includegraphics[scale=0.45]{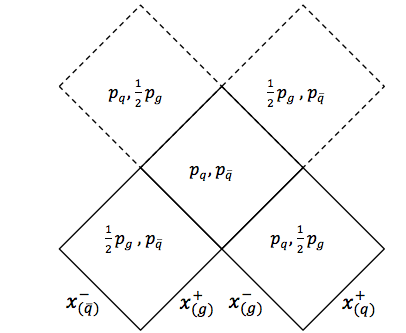}
\caption{The parameter plane picture for the $\q\g\qbar$ system.
The dash lines indicate the turnover regions, normally neglected. }
\label{fig:qgqbarparameterplane}
\end{figure}

Although Fig.~\ref{fig:qqgevolution} is useful to visualize the time
evolution of the system, the parameter plane picture is most convenient
when addressing the kinematics \cite{Sjostrand:1984ic}. This is a
diagram that displays the different string regions in terms of the
light-cone four-vectors defining each region, i.e.\ $p^+_{\q} = p_{\q}$,
$p^-_{\qbar} = p_{\qbar}$ and $p^+_{\g} = p^-_{\g} = p_{\g} / 2$ in the
$\q\g\qbar$ case, whose parameter plane is displayed in Fig.~\ref{fig:qgqbarparameterplane}.
The low regions represent the states in which none of the partons
have lost their energy, corresponding to the two string regions in
view 1 of Fig.~\ref{fig:qqgevolution}, the $\q\g$ and the $\g\qbar$
string pieces. The intermediate region corresponds to the new string
piece created from the $\q$ and $\qbar$ momenta once the gluon has lost
all its energy. Finally, the upper regions are related to the two string
pieces formed when $\g$ re-appears. Although the complete parameter plane
picture (for half a period) is the one shown
in Fig.~\ref{fig:qgqbarparameterplane}, the dashed upper regions are
normally neglected, since the system is assumed to fragment before then.
This reasonable assumption avoids a large number of complications
for handling fragmentation in these regions. The three remaining regions
are then formed by the combination of one $+$ component and one $-$ one,
where $\pm$ no longer relates to motion along the $\pm z$ axis, but
more generically denotes the reference vector directed towards ($+$)
or away from ($-$) the $\q$ end of the system.

From the parameter plane picture, the equations defining the hadron
properties and the fragmentation process of section~\ref{sec:twoparton}
can be easily generalized. For the $\q\g\qbar$ system, the hadron
momentum can generically be written as
\begin{eqnarray}
p_h & = & x^+_{\q} p^+_{\q} + x^-_{\g} p^-_{\g} + x^+_{\g} p^+_{\g}
    + x^-_{\qbar} p^-_{\qbar} ~, \nonumber \\
& = & x^+_{\q} p_{\q} + \frac{1}{2} (x^+_{\g}+ x^-_{\g}) p_{\g}
  + x^-_{\qbar}p_{\qbar} ~.
\end{eqnarray}
The hadron mass enters via the constraint $p_h^2 = m_h^2$. The other
Lorentz invariant variable $\Gamma$ is obtained from the $\hx^{\pm}$
fractions defined in section~\ref{sec:stringbreaking} as
\begin{equation}
\Gamma =  (\hx^+_{\q} p^+_{\q} + \hx^+_{\g} p^+_{\g} + \hx^-_{\g} p^-_{\g}
  + \hx^-_{\qbar} p^-_{\qbar} )^2 ~.
\end{equation}
The level lines of constant $m_h$ and constant $\Gamma$ again give
hyperbolae inside each string region, where physically allowed,
which connect at the borderline between regions. As before there will
be (at most) one allowed solution to a given ($m_h, \Gamma)$ pair,
which can be found by starting in the current region and, if not found
there, step by step move on to other possible regions. There are a
number of complications that have to be overcome to do this
\cite{Sjostrand:1984ic}.

\begin{figure}[tp]
\centering
\includegraphics[scale=0.36]{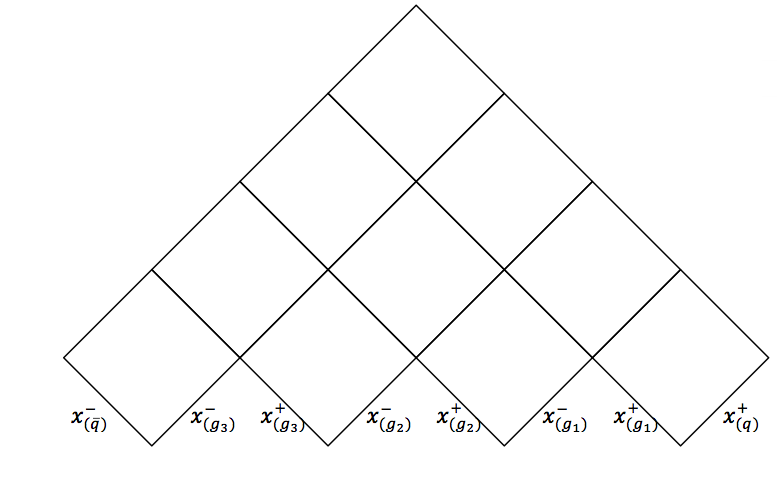}
\caption{The parameter plane picture for a multiparton system composed
by five partons.}
\label{fig:multipartonparameterplane}
\end{figure}

The parameter-plane picture can be extended to a multiparton
system, resulting in the most convenient approach to study the
kinematics of any multiparton system. As an example, the parameter
plane for a system consisting of three gluons, one quark and one
antiquark is depicted in Fig.~\ref{fig:multipartonparameterplane},
where the turnover regions have been ignored. In such a system,
there are four low regions, or initial regions, and six intermediate
regions. The number of initial and intermediate regions can be
generalized for any multiparton system formed by $n$ partons, out of
which $n-2$ are gluons, as $n-1$ initial regions and $(n-1) (n-2) / 2$
intermediate regions. The expression for the hadron four-momentum can
also be generalized to an $n$-parton system by accounting for the
momenta taken from each parton as
\begin{equation}
p_h = x_{\q}^+ p_{\q}^+ + x_{\qbar}^- p_{\qbar}^-
+ \sum_{i=1}^{n-2} (x_{g_i}^+ p^+_{g_i} + x_{g_i}^- p^-_{g_i}) ~,
\end{equation}
where usually most of the $x^{\pm}$ vanish. Apart from these aspects,
the rest of the properties are similarly determined as in previous cases.

\subsection{Fragmentation implementation summary}
\label{sec:lundimplementation}

The fragmentation process in \textsc{Pythia} is based on the four-momenta
of the partons created in the (semi)perturbative stages of the collision
process, plus the partons in the beam remnants \cite{Sjostrand:2004pf}.
By the colour-connection between those partons, an LHC event is likely to
contain several $\q\g_1\g_2\ldots\g_{n-2}\qbar$ systems, that can be
handled separately.

The production of each new hadron begins with the selection at random of
whether to split it off from the $\q$ end or from the $\qbar$ one of the
system. The flavour of a new $\q\qbar$ break of the string (where $\q$ may
also represent an antidiquark), leads to the formation of a new hadron,
as already described. Its mass is selected, according to a Breit--Wigner
for short-lived particles with a non-negligible width. The transverse
momentum is obtained as the vector sum of those of the hadron constituents,
assuming that the old and new breakup vertices are in the same region.
Then the longitudinal momentum fraction $z$ is picked according to
the probability distribution in  eq.~(\ref{eq:probabilityofz}), with
the difference that the hadron mass has to be replaced by the transverse
ditto, $m_{h} \to m_{\perp h}$. In a simple $\q\qbar$ system, the new
breakup vertex is easily obtained from the $(m_{\perp h},z)$ pair.
Else the $\Gamma_i$ value of the new breakup is calculated using
eq.~(\ref{eq:newoldgammas}), again with $m_{h} \to m_{\perp h}$, and a
solution is sought to the  $(m_{\perp h}, \Gamma_i)$ pair of equations.
Vertex $i$ may end up in the same string region as $i - 1$, or involve
a search in other regions. Among technical complications of this search
is that the transverse directions are local to each string region,
which leads to discontinuities in the hyperbolae of constant
$m_{\perp h}$ at the borders between string regions, that would not be
there for $\pT = 0$.

The random steps from both string ends continue until the remaining
invariant mass of the system is deemed so small that only two final
hadrons should be produced. Details on this final step can be found in
Appendix~\ref{sec:spacetimefinal}, along with the challenges encountered
when implementing the space--time picture and the methods applied to
solve them. Had the fragmentation always proceeded from the $\q$ end,
say, the final step would always have been at the $\qbar$ end, with
the minor blemishes of this step concentrated there. Now these are
instead smeared out over the whole event.

\section{The space--time description}
\label{sec:spacetime}

So far, the fragmentation process in \textsc{Pythia} was developed
in terms of the energy--momentum fractions $x^{\pm}$ and $z^{\pm}$
of breakup vertices and hadrons, presented in
section~\ref{sec:stringbreaking}. Therefore, the location of the breakup
vertices is only specified in the energy--momentum picture. In order to
study the density of hadron production, this information should first
be translated to the space--time one, which will be done in this section.

\subsection{The two-parton system}
\label{sec:breakuppoints}

To begin, consider a breakup point $i$ in a simple $\q\qbar$ system.
Its location with respect to the origin of the energy--momentum
picture, where $\q$ and $\qbar$ have been created, is given by the
$\hx^{\pm}$ fractions. Then, considering $p^+$ to be the $\q$
four-momentum and $p^-$ the $\qbar$ four-momentum, the location
of breakup $i$ in the energy--momentum picture is defined as
$\hx_i^+ p^+ + \hx_i^- p^-$. Recalling the linear relation
between space--time and energy--momentum, eq.~(\ref{eq:xplinearity}),
the space--time location of breakup point $i$ thereby is defined as
\begin{equation}
v_i = \frac{\hx_i^+ p^+ + \hx_i^- p^-}{\kappa} ~.
\label{eq:spacetimebreakup}
\end{equation}
Note that the $\q_i$ and $\qbar_i$ generated by the string break are
considered to be created in the same space--time vertex, even when
quark masses and transverse momenta are included, such that  $\q$ and
$\qbar$ have to tunnel some distance apart before becoming on-shell.
Such effective vertices in practice is the best one can do.

\begin{figure}[tp]
\centering
\includegraphics[scale=0.35]{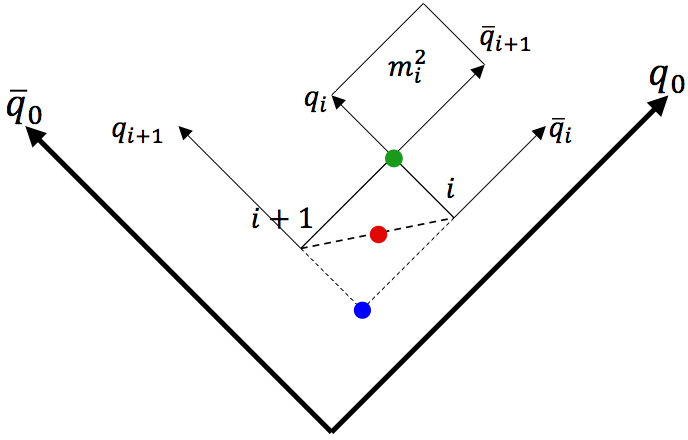}
\caption{Hadron formation in a $\q\qbar$ system. The blue, red
and green dots represent the ``early'', ``middle'' and ``late''
definitions of hadron production points, respectively.}
\label{fig:hadprodpoint}
\end{figure}

Since hadrons are formed by two adjacent string breaks, the hadron
production point should be related to these two. But the definition
cannot be unique, since hadrons are composite and extended particles.
For that reason, we propose three alternative definitions, illustrated
in Fig.~\ref{fig:hadprodpoint}. Two breakup points, $i$ and $i+1$,
with space--time coordinates  $v_i$ and $v_{i+1}$, together define
the $\q_i\qbar_{i+1}$ subsystem that forms hadron $i$. One obvious
choice is then to define the hadron production point as the average
of the two,
\begin{equation}
v^h_i = \frac{v_i + v_{i+1}}{2} ~,
\label{eq:middle}
\end{equation}
red dot in the figure. Alternatively to this ``middle'' definition, the
``late'' hadron production point is where the two partons forming the
hadron cross for the first time, green dot. Taking into account the
hadron four-momentum $p_h$, the ``late'' hadron production point is offset
from the ``middle'' definition as
\begin{equation}
v^h_{l,i} = \frac{v_i + v_{i+1}}{2} + \frac{p_h}{2\kappa} ~.
\label{eq:late}
\end{equation}
Finally, an ``early'' definition, blue dot,
is given by
\begin{equation}
v^h_{e,i} = \frac{v_i + v_{i+1}}{2} - \frac{p_h}{2\kappa},
\label{eq:early}
\end{equation}
which is where the backwards light cones of the $\qbar_i$ and $\q_{i+1}$
vertices cross, just like the ``late'' definition is where the forwards
light cones cross. In a causal world, this would be the latest time for
information to be sent out that can correlate the breakup vertices
to give the correct hadron mass. Note that the two endpoint hadrons are
situated on the light cone with this ``early'' definition. The different
results obtained with the three alternative definitions can be used as a
measure of uncertainty, see section~\ref{sec:ltdist}. If not stated
otherwise, the choice in this article is the ``middle'' definition of
eq.~(\ref{eq:middle}).

\subsection{More complex topologies}
\label{sec:complextopologies}

\begin{figure}[tp]
\centering
\includegraphics[scale=0.25]{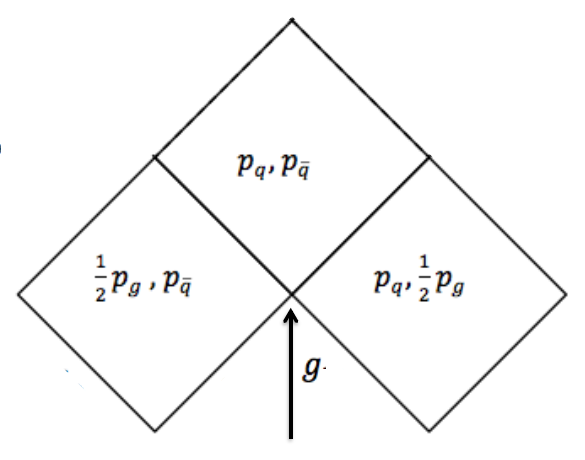}
\captionsetup{justification=centering}
\caption{Parameter plane for a $\q\g\qbar$ system.}
\label{fig:offsetoneg}
\end{figure}

Multiparton systems are more complicated to address than a single
$\q\qbar$ string, as already demonstrated for the energy--momentum picture.
Their complexity also affects the space--time implementation, which has to
be extended to include several string regions. Initially consider a
$\q\g\qbar$ system formed by massless partons, with a parameter plane as
in Fig.~\ref{fig:offsetoneg}, ignoring the turnover regions. Since each
string region separately behaves like a  simple $\q\qbar$ system,
eq.~(\ref{eq:spacetimebreakup}) can be used. Nevertheless, the
intermediate region is formed after the gluon has lost all of its energy,
at a different location in space--time than the initial regions, so an
offset has to be taken into account for it. From the linear relation between
space--time and energy--momentum, the space--time offset for this region can be
calculated as $v_{\mathrm{reg}} = p_{\g}/2\kappa$, where $p_{\g}$ is the
four-momentum of the gluon. The factor of $1/2$ accounts for the fact
that a gluon loses four-momentum twice as fast as a $\q$ or $\qbar$,
since it transfers four-momentum to two string pieces. Thus, the
space--time location of a breakup located in the intermediate region is
given by
\begin{equation}
v_i = \frac{\hx_i^+ p^+ + \hx_i^- p^-}{\kappa} + \frac{p_{\g}}{2\kappa} ~.
\label{eq:offsetonegluon}
\end{equation}

\begin{figure}[tp]
\centering
\includegraphics[scale=0.3]{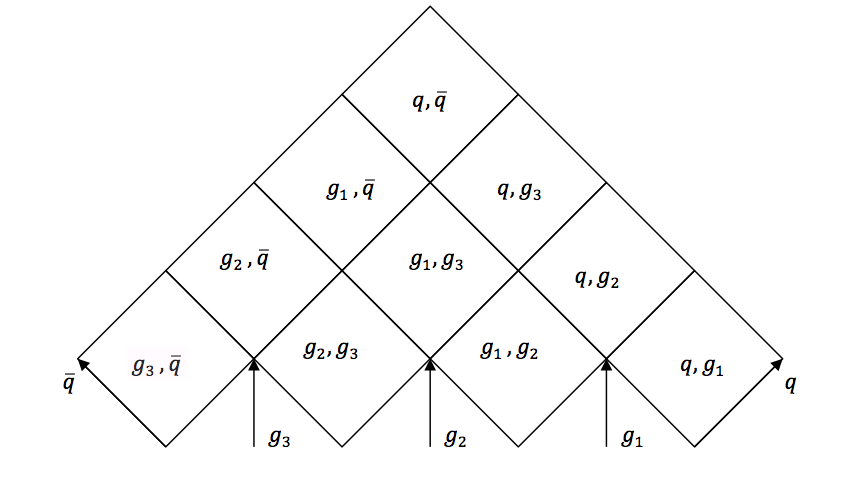}
\captionsetup{justification=centering}
\caption{Parameter plane for a five parton system.}
\label{fig:offset}
\end{figure}

If the system is composed of more than one gluon, also more than
one intermediate region has to be taken into account, as illustrated
in Fig.~\ref{fig:offset}. In such cases, more gluons have to be
included when determining the space--time offset of some intermediate
regions, such as the $\q\g_3$ one. This region is created when both
$\g_1$ and $\g_2$ have lost their energies, giving an offset
$v_{\mathrm{reg}} = (p_{g_1}+p_{g_2}) /2\kappa$.

A general expression for the space--time offset of any intermediate
region in any multiparton system can be easily defined, if all partons
are numbered consecutively, starting from the $\q$ end, and region
labels $jk$ are for ones containing four-momenta from partons $j$ and
$k, k\geq j$. The $jk$ region offset is found to be
\begin{equation}
v_{jk} =\sum_{m = j + 1}^{k - 1} \frac{p_m}{2\kappa} ~,
\end{equation}
where $p_m$ is the four-momentum of parton $m$, and for a breakup vertex
in this region it thus holds that
\begin{equation}
v_i = \frac{\hx_i^+ p^+_j + \hx_i^-p^-_k}{\kappa}
+ \sum_{m = j + 1}^{k - 1}\frac{p_m}{2\kappa}.
\end{equation}
While seemingly simple enough, there are a number of significant challenges
to a robust implementation in a multiparton configuration, in part
paralleling similar problems in the energy--momentum picture
\cite{Sjostrand:1984ic}, in part going further. There are two main problems:
the determination of the space--time location of the final breakup in
the system, and the non-physical values of the $\hx^{\pm}$ fractions that
can arise when fragmentation moves to a new region. Those issues are further
explained in appendices \ref{sec:spacetimefinal} and \ref{sec:nonphysical},
respectively, along with the solutions found to properly implement the
space--time picture.

\subsection{Gluon loops}

\begin{figure}[tp]
\centering
\subfloat[String configuration]{
    \includegraphics[scale=0.34]{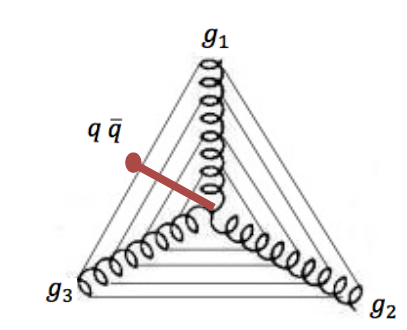}}
\hspace{2mm}
\subfloat[Parameter plane]{
    \includegraphics[scale=0.28]{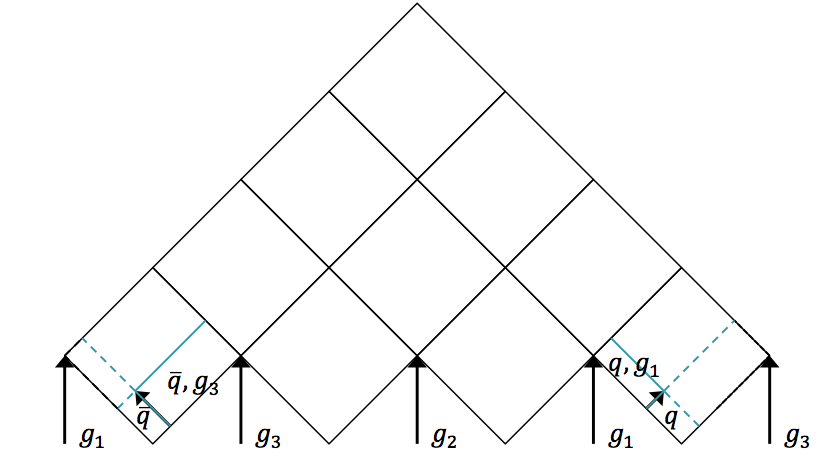}}
\caption{The string configuration and the corresponding parameter plane
for a three gluon-loop topology. In the two initial endpoint regions the
full lines indicate the ``active area'', and the dashed ones the
complementary excluded one.}
\label{fig:gluonloop}
\end{figure}

So far, gluons have only appeared in open strings between a $\q$ and a
$\qbar$ end, but it is also possible to have closed gluon loops,
as exemplified in Fig.~\ref{fig:gluonloop}a for a $\g\g\g$ system.
In order to reduce the problem to a familiar one, an initial $\q\qbar$
is generated by string breaking in one of the string regions. This
break should be representative of what ordinary fragmentation is
expected to give. Thus the region is chosen at random, but with a bias
towards ones with larger masses, where more ordinary string breaks are
to be expected. Inside that region, the $\Gamma$ value of the vertex
is chosen according to eq.~(\ref{eq:probabilityofgamma}), and a further
random choice gives the longitudinal location of the breakup. Having
taken this step, the $n$-gluon-loop topology is effectively mapped onto
an $(n+1)$-parton open string with $\q$ and $\qbar$ as endpoints.
The key difference is that, unlike open strings considered so far,
$\Gamma_{\q} = \Gamma_{\qbar} \neq 0$.

As an example, the parameter plane for a gluon-loop consisting
of three gluons is displayed in Fig.~\ref{fig:gluonloop}b.
In this case, the string between $\g_1$ and $\g_3$ has broken into two
string pieces, generating two new string regions, $\g_1\q$ and $\g_3\qbar$.
Although the full $\g_1\g_3$ region is duplicated in the parameter plane,
in the right endpoint region only the ``active area'' between $\q$ and
$\g_1$ is open to fragmentation, while the left endpoint region only
uses the complementary area between $\g_3$ and $\qbar$. Apart from that,
the fragmentation process can now play out in the same way as for an open
string, with the same rules for the space--time locations of the breakups.
Note that the $\q$ and $\qbar$ ``endpoints'' correctly will be assigned the
same creation vertex in this procedure.

\subsection{Smearing in transverse space}
\label{sec:smearing}

Strings can be viewed as the center of cylindrical tubelike regions
of directed colour flow. So far we have assigned production vertices
as if they all were in the very center of the string. A more realistic
picture is to introduce some transverse smearing. For simplicity this
is done according to a two-dimensional Gaussian
\begin{equation}
f (x,y)  \propto \exp \left(-\frac{x^2+y^2}{2\sigma^2} \right) ~,
\end{equation}
where $x$ and $y$ are transverse spatial coordinates and $\sigma$ is the
width of the distribution.

The width of the string should be of typical hadronic scales, but related
to confinement in two dimensions rather than three. Taking the proton
radius $r_{\p}\approx 0.87$~fm \cite{Patrignani:2016xqp} as starting point,
the default $\sigma = r_{\p}/\sqrt{3}$ then gives a
\begin{equation}
r_{\perp,\p}^2 = \langle x^2+y^2 \rangle = 2 \sigma^2 = \frac{2}{3} r_{\p}^2.
\end{equation}

The smearing in transverse space might generate unwanted situations,
such as negative values for the $\Gamma$ parameter of the breakup points.
Since the space--time location is first obtained from the fragmentation
picture in the longitudinal direction, the squared invariant time should
not change when introducing smearing. Therefore, the time coordinate is
adjusted after including the smearing in transverse space, in order to
retain the $\Gamma$ value determined by the longitudinal scheme.
Alternative procedures could be envisioned, in particular when the
collision process itself does not happen in the origin, but for now
this smearing possibility is good enough to indicate trends.

\subsection{Massive quark implementation}
\label{sec:massiveimplementation}

As illustrated in section~\ref{sec:massivequarks}, the origin of the
massive and massless oscillations are displaced for technical reasons;
correspondingly the initial point of the massive oscillation is offset
from the origin of the space--time coordinate system. Since the
fragmentation process is performed in the massless system, the
space--time locations of the breakups have to be adjusted.

\begin{figure}[tp]
\centering
\subfloat[Massive offset calculation]{
    \includegraphics[scale=0.35]{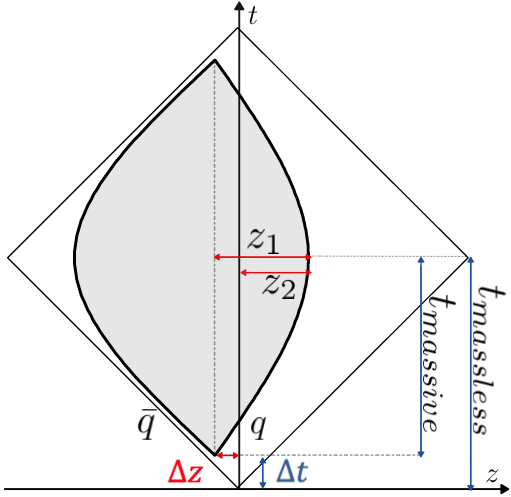}} \hspace{5mm}
\subfloat[Endpoint correction]{
    \includegraphics[scale=0.35]{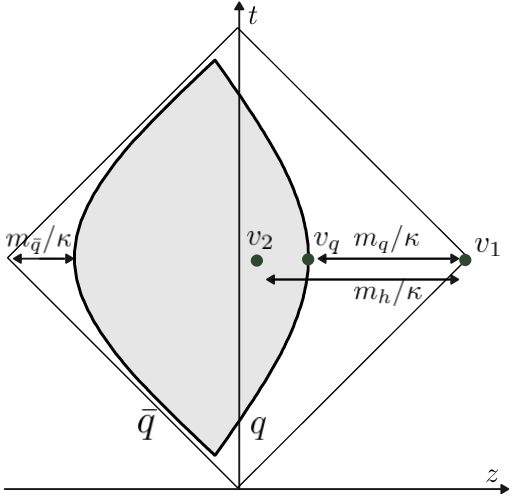}}
\caption{Massive $\q\qbar$ system and equivalent massless
system. The grey area corresponds to the physical region.}
\label{fig:massiveoffset}
\end{figure}

To determine the offset, consider the $\q\qbar$ system in
Fig.~\ref{fig:massiveoffset}a, studied in the CM frame, with
$\q/\qbar$ moving in the $\pm z$ directions. In this case
the $\q$ and $\qbar$ masses are different, with $m_{\q} > m_{\qbar}$.
At time $t=0$ we have $p_0 = p_{z,\q} = -p_{z,\qbar}$ and
$\ECM = E_{\q} + E_{\qbar}$, with $p_0$, $E_{\q}$ and $E_{\qbar}$
given by standard two-body decay kinematics.
The massive oscillation in Fig.~\ref{fig:massiveoffset}a is offset both
in time and $z$-component of space, represented as $\Delta t$ and
$\Delta z$. The former can be determined from the difference
between the time coordinates at which the massless and massive quarks
lose their three--momenta, $t_{\mathrm{massless}}$ and $t_{\mathrm{massive}}$
in Fig.~\ref{fig:massiveoffset}a, i.e.\,
\begin{equation}
\Delta t = t_{\mathrm{massless}} - t_{\mathrm{massive}}
= \frac{\ECM}{2\kappa} - \frac{p_0}{\kappa}
= \frac{\ECM - 2p_0}{2\kappa} ~.
\label{eq:timeoffsetmassive}
\end{equation}
The process to define the space offset is slightly different. In
Fig.~\ref{fig:massiveoffset}a, the distance of the massive $\q$ endpoint
to the centre of the massive oscillation is denoted $z_1$, while $z_2$ is
the distance of the massive $\q$ endpoint to the centre of the massless
system. The equation of motion then gives
\begin{equation}
\Delta z = z_1 - z_2 = \frac{1}{\kappa} \left( \frac{\ECM}{2} - m_{\q} \right)
- \frac{1}{\kappa} (E_{\q}-m_{\q}) = \frac{E_{\qbar} - E_{\q}}{2\kappa} ~.
\label{eq:zoffsetmassive}
\end{equation}
The time and space offsets can be combined as
\begin{equation}
v_{\mathrm{offset}} = \frac{1}{\kappa\ECM}
\left( (E_{\qbar} - p_0) p_{0\q} + (E_{\q} - p_0) p_{0\qbar} \right) ~,
\label{eq:massiveoffsetvectorial}
\end{equation}
where $p_{0\q}$ and $p_{0\qbar}$ are the four-vectors of the equivalent
massless system, cf.\ eq.~(\ref{eq:masslessrefvec}).
Hence, for each vertex in a region formed by at least one massive quark,
the space--time location is defined as usual from the massless system,
$v_{\mathrm{massless}}$, and then corrected to
\begin{equation}
v_{\mathrm{correct}} = v_{\mathrm{massless}} - v_{\mathrm{offset}} ~.
\label{eq:massiveoffsetcorr}
\end{equation}
For more complex topologies, such as multiparton systems consisting of
a massive $\q$ and/or $\qbar$ and several gluons, the effect of the
massive $\q$ or $\qbar$ is only non-negligible in the lowest respective
endpoint region. Therefore, the massive correction is only performed in those regions.

Also the space--time location of the massive endpoint quark ``vertex'' has
to be offset, away from what it would have been for a massless
quark. This is exemplified in Fig.~\ref{fig:massiveoffset}b, for the
same massive $\q\qbar$ system as before. The three vertices $v_1$,
$v_{\q}$ and $v_2$  correspond to the space--time location of the massless
endpoint, the massive turning point and the closest breakup to the
massless endpoint, respectively. The system can be studied in a Lorentz
frame where the three vertices are simultaneous,
$v_{1,t} = v_{2,t} = v_{\q,t}$. Then, linearity between energy--momentum
and space--time gives
\begin{equation}
\begin{gathered}
v_{1,z} - v_{2,z} = \frac{m_h}{\kappa} ~,\\
v_{1,z} - v_{\q,z} = \frac{m_{\q}}{\kappa} ~,
\end{gathered}
\label{eq:verticessystemmassive}
\end{equation}
where $m_{\q}$ is the mass of the heavy quark and $m_h$
the mass of the hadron formed from the vertices $v_1$ and
$v_2$. From this $v_{z,\q}$ can be extracted. Recast in Lorentz-invariant
four-vector notation, this gives
\begin{equation}
v_{\q} = v_1 + \frac{m_{\q}}{m_h} (v_2 - v_1)  ~.
\label{eq:finalendpointmassivecorr}
\end{equation}
Note that, after the correct endpoint location has been determined,
the offset correction of eq.~(\ref{eq:massiveoffsetcorr}) has to be
included. If a system is formed by a  massless $\q$ and a massive
$\qbar$, say, eq.~(\ref{eq:finalendpointmassivecorr}) has to be applied
only to the massive $\qbar$, while the offset in
eq.~(\ref{eq:massiveoffsetcorr}) has to be used both for $\q$ and $\qbar$.

A final feature is that the oscillation period for a hadron composed
of massive quarks is shorter than a same-mass one with massless quarks.
This discrepancy only affects the estimation of the ``late'', $v_{l}^h$,
and ``early'', $v_{e}^h$, definitions of hadron production points.
The expression in eq.~(\ref{eq:late}) and eq.(~\ref{eq:early}) now become
\begin{equation}
v_{l/e}^h = \frac{v_i + v_{i+1}}{2} \pm
\alpha_{\mathrm{red}} \, \frac{p_h}{2\kappa} ~,
\end{equation}
where $\alpha_{\mathrm{red}}$ accounts for the reduced oscillation period.
This parameter is determined in the hadron rest frame by the absolute
three-momentum of the quarks forming the hadron
\begin{equation}
\alpha_{\mathrm{red}} = \frac{p_0}{m_h} = \frac{\sqrt{ (m_h^2 - m_{\q}^2
- m_{\qbar}^2)^2 - 4m_{\q}^2 m_{\qbar}^2}}{m_h^2} ~,
\end{equation}
where $m_h$ is the mass of the hadron and $m_{\q}$ and $m_{\qbar}$
the masses of the constituent $\q$ and $\qbar$, respectively.
Needless to say, these semiclassical estimates of oscillation periods
cannot be taken too literally. It could be argued that all hadrons,
light as heavy, have hadronic sizes of order 1~fm, and should
have essentially common oscillation periods related to that. That would
give us problems notably for pions, however, which are abnormally
light in relation to their size.

\subsection{Other implementation details}
\label{sec:junctionsetc}

Up until now, only open $\q\g_1\g_2\ldots\g_{n-2}\qbar$ and closed
$\g_1\g_2\ldots\g_n$ strings have been considered. A third possibility is
junction topologies, wherein three string pieces meet in a common vertex
\cite{Sjostrand:2002ip}, and whereby the junction effectively carries
the baryon number of the system. Such topologies can arise e.g.\ when
the three valence quarks are all kicked out of an incoming proton,
but there are also scenarios in which further junctions and
antijunctions may be formed \cite{Christiansen:2015yqa}.

A junction system consists of three different ``legs'', each stretched
from an endpoint quark via a number of gluons in to the junction.
In \textsc{Pythia} the fragmentation process is most conveniently
defined in the rest frame of the junction. Here the total energy of
each leg is determined, and the two legs with the lowest energies are
fragmented from the respective $\q$ end inwards. The process stops when
the next step would require more energy than left in the leg.
Once the two initial legs have fragmented, the two leftover $\q$
from the respective last breaks are combined to create a diquark.
Together with the third leg and its original endpoint $\q$, this
diquark defines a final string system, which now fragments as a
normal open string.

The assignment of space--time locations in junction topologies
introduces no new principles, but requires some extra bookkeeping.
The three junction legs are considered as three different systems,
to be dealt with in the same order as they fragmented, starting from
the leg with the lowest energy.

Low-invariant-mass systems hadronize about as high-mass ones, even if
kinematics is more constrained. The exception is when the invariant
mass of the system is so low that only one hadron can be formed.
In such cases, the ``early'' hadron production point is at the origin
of the $\q\qbar$ system, i.e.\ $v_e^h= (0; 0, 0, 0)$.
Note that smearing in transverse space will give rise to negative
squared invariant times in such cases. This is not a problem if
the reason is that the collision of two Lorentz-contracted proton
``pancakes'' naturally would lead to a spread of $x,y$ coordinates
of collisions at $t=0$. The ``middle'' and ``late'' definitions are
calculated from the four-momentum $p_h$ of the hadron as
$v^h = p_h/2\kappa$ and $v_l^h = p_h/\kappa$, respectively.

Finally, many of the hadrons produced during the string fragmentation
are unstable and decay further, a process known as secondary particle production. In such
cases the invariant lifetime is selected at random according to an
exponential decay, $P(\tau) \propto \exp(-\tau / \langle \tau \rangle)$,
where $\langle \tau \rangle$ is the tabulated average lifetime
\cite{Patrignani:2016xqp}. For short-lived particles it is rather
the width $\Gamma$ of the mass distribution that is known, and
then one can use $\langle \tau \rangle = \hbar c / \Gamma$.
Given a known hadron production vertex, the decay one becomes
\begin{equation}
v_{\mathrm{decay}} = v_{\mathrm{production}} + \tau \, \frac{p_h}{m_h} ~,
\end{equation}
for a hadron with four-momentum $p_h$ and mass $m_h$. This equation
can be used recursively through decay chains, also e.g.\ for leptons.

Truly stable particles are only $\e^{\pm}$, $\p$, $\pbar$, $\gamma$
and the neutrinos. Also some weakly decaying particles with long
lifetimes are effectively treated as stable by default: $\mu^{\pm}$,
$\pi^{\pm}$, $\K^{\pm}$, $\K_L^0$ and $\n/\nbar$.

\subsection{A comparison of time scales}
\label{sec:pertpict}

In this article we only address the space--time picture of hadronization.
In the context of a hard collision process, say $\q\g \to \q\g$, also 
perturbative emission of further partons off the two scattered partons 
is extended in space and time. This is related to the regeneration time 
of the QCD field, mainly consisting of gluons, at typical time scales 
of order
\begin{equation}
t_{\mathrm{regen}} \sim \frac{\hbar c \, E}{\pT^2} 
= \frac{\hbar c}{\pT} \, \frac{E}{\pT} \sim \tau_{\mathrm{regen}} \, \gamma 
\label{eq:formtime}
\end{equation}
for emitted partons of energy $E$ and transverse momentum $\pT$
\cite{Dokshitzer:1991wu}. This expression is conveniently split into 
a ``Heisenberg uncertainty'' factor ($\pT$ is a measure of the 
off-shellness of intermediate propagators) and a ``time dilation'' 
factor, as indicated. Similar relations hold for emissions off the 
two incoming partons. 

Typically, parton shower descriptions in event generators such as 
\textsc{Pythia} stop at scales of the order 
$\p_{\perp\mathrm{min}} = 0.5 - 1$~GeV, mainly because $\as$ becomes so 
big that perturbation theory cannot be trusted below that. (The current
default value for \textsc{Pythia} final-state radiation is 0.5~GeV,
but that is the $\pT$ for each daughter of a branching with respect 
to the mother direction, meaning a separation of 1~GeV between the
two daughters. eq.~(\ref{eq:formtime}) should not be trusted up to 
factors of 2 anyway.)  This corresponds to a 
$\tau_{\mathrm{regen}} \approx 0.25$~fm, say, to be compared with the average 
hadronization time $\langle \tau_{\mathrm{had}} \rangle \approx 1.3$~fm 
(see section~\ref{sec:timeevolution} below), i.e. about a factor five 
difference. To a good first approximation, the simulated perturbative activity 
can therefore be viewed as happening in a single point as far as the 
hadronization process is concerned. This is even more so for the hard
perturbative activity that gives rise to separate jets, for which 
$\pT \gg 1$~GeV. The emissions that possibly are simulated below 1~GeV
can only give small wrinkles on the strings stretched between the main 
partons.

The comparison of invariant time generalizes to hold everywhere in an 
event, since time dilation works the same way for showers and hadronization.
That is, a perturbative splitting at high energy and low $\pT$ may 
occur at large time scales as measured in the rest frame of the event, 
when hadronization already started in the central region, 
but still well before it will begin in the part of the event that
could be affected by the splitting.

At the end of the \textsc{Pythia} showers, the total number of partons 
in a typical LHC event is roughly half of the number of primary hadrons
later produced. Given that the size, in each of three spatial dimensions,
is only a fifth for the partonic system compared with the hadronic one, 
it might seem that the partonic density is much higher than the the 
hadronic one, and that partonic close-packing would be a more severe 
issue than hadronic ditto. Partons don't have a well-defined size, 
however. A newly created parton could be assigned a vanishingly small 
size, and then the colour field surrounding it would expand with the 
speed of light. Thus the partonic size could be equated with the time 
since creation, multiplied by a standard time dilation factor.
 
At early times the partonic system of a collision therefore expands 
in size at about the same rate as the size of partons, and any net effect 
comes from the rise of the total number of partons as the cascade 
evolves from early times. Here the colour coherence phenomenon enters, 
however \cite{Dokshitzer:1991wu}. It is the obervation that the two 
daughters of a $ \q \to \q\g$ or $\g \to \g\g$ branching share a 
newly-created colour-anticolour pair, that cannot contribute to the 
radiation until the partons are more separated than the wavelength 
of the further radiated partons. This gives a mechanism for close-packing 
avoidance, in event generators implemented in terms of angular or $\pT$ 
ordering of radiation.   

Had the parton shower been allowed to evolve further than the current
cutoff, the partonic multiplicity and the partonic overlap would have 
increased as the $\Lambda_{\mathrm{QCD}}$ scale of $\approx 0.3$~GeV is 
approached. By then the naive size of partons would be of the order of 
0.7~fm, which is about the expected transverse size of strings, and
soft partons emitted at this stage form part of the emergent strings. 
We don't know how to model these late stages of the cascade, but any
effects coming from them are included in the tuned parameters of the 
string fragmentation framework.  

The picture painted here is based on studying one partonic cascade.
Since protons are composite object, however, several partonic 
subcollisions can occur when two protons collide --- MPIs. One 
therefore also should worry about the overlap of cascades from 
different MPIs --- partonic rescattering. In part this issue has 
been studied \cite{Corke:2009tk}, and shown to give small effects. 
That study only included the effects of parton multiplication by
initial-state radiation, as encoded in parton distribution functions,
and thus did not address the effects of collisions between partons 
from two separate MPI subcollisions. In general, however, MPIs
occur at different transverse locations when the two Lorentz-contracted
protons collide, and the products move out in different rapidities and 
azimuthal directions. Also here it is therefore plausible with only
minor overlap at early times and large perturbative scales. (In a 
relative sense; most MPIs do not have all that large $\pT$ values.) 
The overlap becomes relevant at later scales, where colour reconnection
is the currently favoured mechanism for interactions between the 
emerging colour fields.

Another issue that we would like to comment on is the folklore that 
``fast particles are produced early''. This would seem to be in 
contradiction with the string picture, where hadronization begins with 
slow particles in the central region and then spreads outwards to
faster particles at later times, roughly along a hyperbola of constant
invariant time. But it is all a matter of what comparison one has in mind,
and what production time definition is used \cite{Bialas:1986cf}.
Consider the ``first'' (``leading'') hadron, i.e.\ the one closest to 
the quark end of a $\q\qbar$ string. For it $\Gamma_{i - 1} = \Gamma_0 = 0$, 
such that eq.~(\ref{eq:newoldgammas}) and eq.~(\ref{eq:gamma}) together give 
\begin{equation}
\Gamma_1 = \frac{1 - z_1}{z_1} \, m_1^2 = (\kappa \tau_1)^2~.   
\end{equation}
The faster the hadron, the earlier the string break in invariant time:
$\Gamma_1 \to 0$ for $z_1 \to 1$. Also the time in the string rest frame,
\begin{equation}
\kappa t_1 = E_{\q} (1 - z_1) + \frac{m_1^2}{4 E_{\q} z_1}~,
\end{equation}
with $E_{\q}$ the quark energy, is decreasing for increasing $z_1$.

This reasoning generalizes: an event with few, fast particles can 
only be obtained when the $\Gamma$ values and the breakup times are 
small. Conversely, events with high multiplicities of lower-momentum
hadrons require high $\Gamma$ values and late hadronization times. 
Whether early or late invariant times, however, the hadronization will 
still start in the middle and spread outwards.

\section{Hadron density studies}
\label{sec:hadrondensity}

We now proceed to study the implications of the model presented
so far. Toy studies are reported for a simple $\q\qbar$ string,
but most results are for $\p\p$ collisions at $\sqrt{s} = 13$~TeV,
for inclusive inelastic nondiffractive events.
Although the $\p\p$ modelling is not yet complete, enough is in
place to perform some first semi-realistic studies that form
the basis for future development. Notably we will estimate the
hadronic density in a few different ways, as a means to highlight
the close-packing of hadrons and the need to consider the consequences
of that.

\subsection{Longitudinal and transverse distributions}
\label{sec:ltdist}

\begin{figure}[tp]
\centering
\includegraphics[scale=0.45]{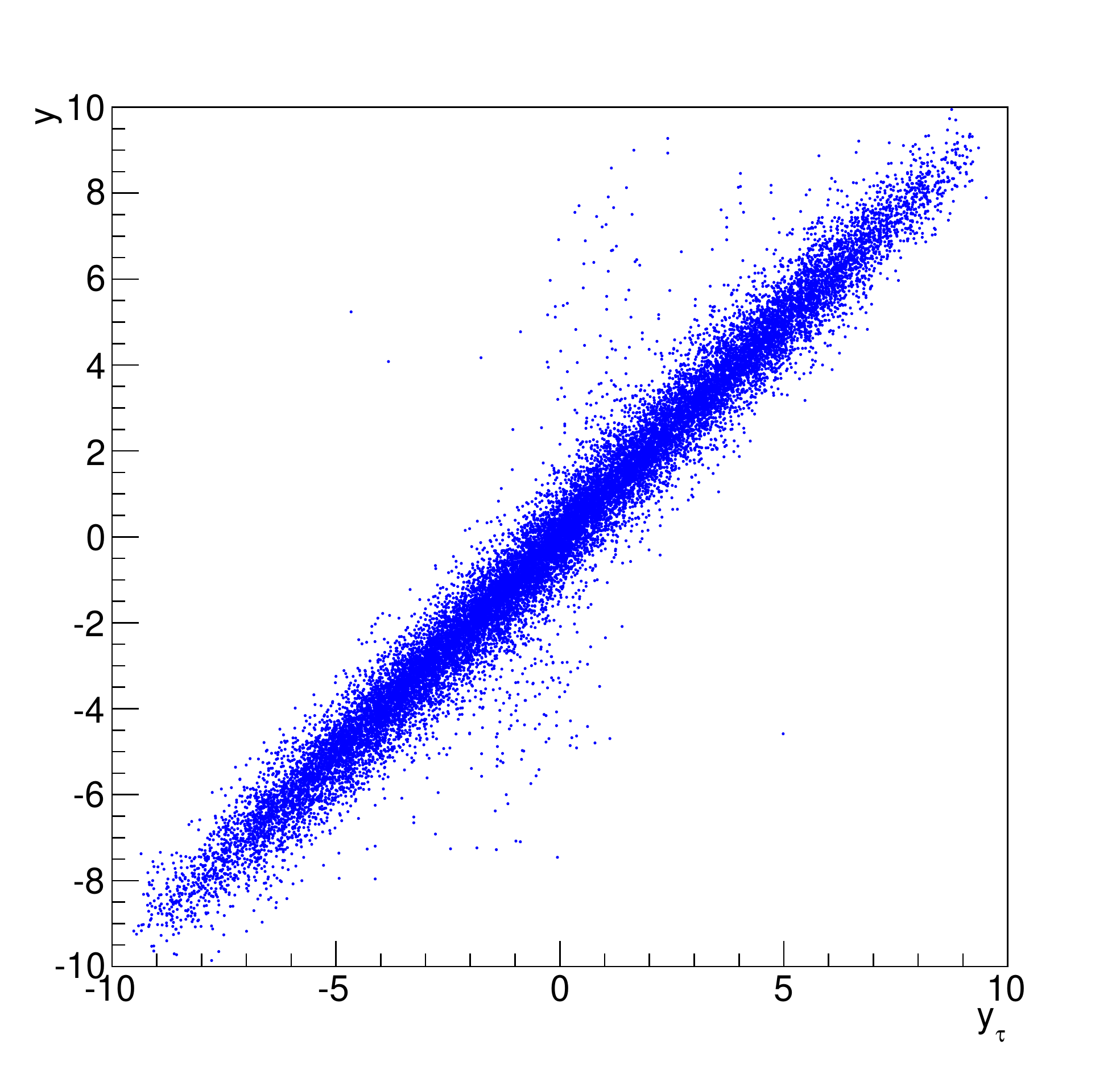}
\caption{Correlation between rapidity, $y$, and the equivalent
space--time rapidity, $y_{\tau}$, for all hadrons in 100
inelastic nondiffractive  $\p\p$ events at $\sqrt{s}=13$~TeV. }
\label{fig:yyTauscatter}
\end{figure}

Three definitions of hadron production points were presented in
section~\ref{sec:breakuppoints}, to allow estimates of the uncertainty
in the description. Here the three resulting longitudinal and transverse
space--time distributions are compared. For the former $y_{\tau}$ is
introduced as a space--time correspondent to ordinary rapidity $y$:
\begin{equation}
y = \frac{1}{2} \log\left( \frac{E + p_z}{E - p_z} \right)
\longrightarrow
y_{\tau} = \frac{1}{2} \log\left( \frac{t + z}{t - z} \right) ~,
\end{equation}
while the latter is shown as a function of $r = \sqrt{x^2 + y^2}$.
Note that the longitudinal variable is dimensionless while
the transverse one is expressed in units of fermi (fm). Although
formally unrelated, the dynamics of string fragmentation introduces
a strong correlation between $y$ and $y_{\tau}$, as illustrated in
Fig.~\ref{fig:yyTauscatter} for the default ``middle'' definition of
production points. The spread from the diagonal comes from a number of
effects, such as the probabilistic fragmentation process, given by
eq.~(\ref{eq:probabilityofz}), and hadronic decays.

\begin{figure}[tp]
\centering
\includegraphics[scale=0.69]{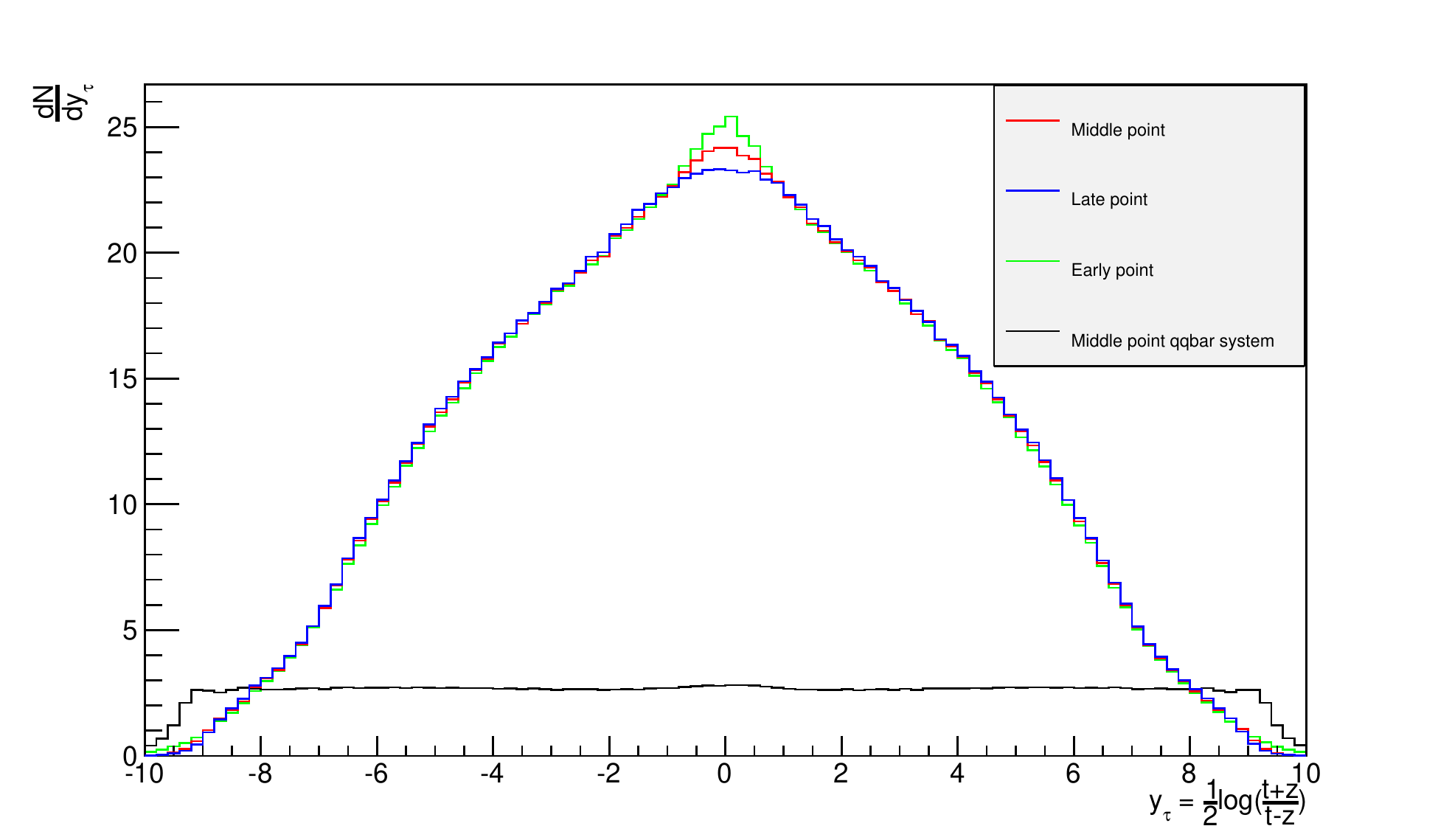}
\captionsetup{justification=centering}
\caption{Longitudinal spectra for $\p\p$ events and
$\q\qbar$ systems, both at $\sqrt{s}=13$~TeV.}
\label{fig:longitudinal}
\end{figure}

\begin{figure}[tp]
\centering
\includegraphics[scale=0.69]{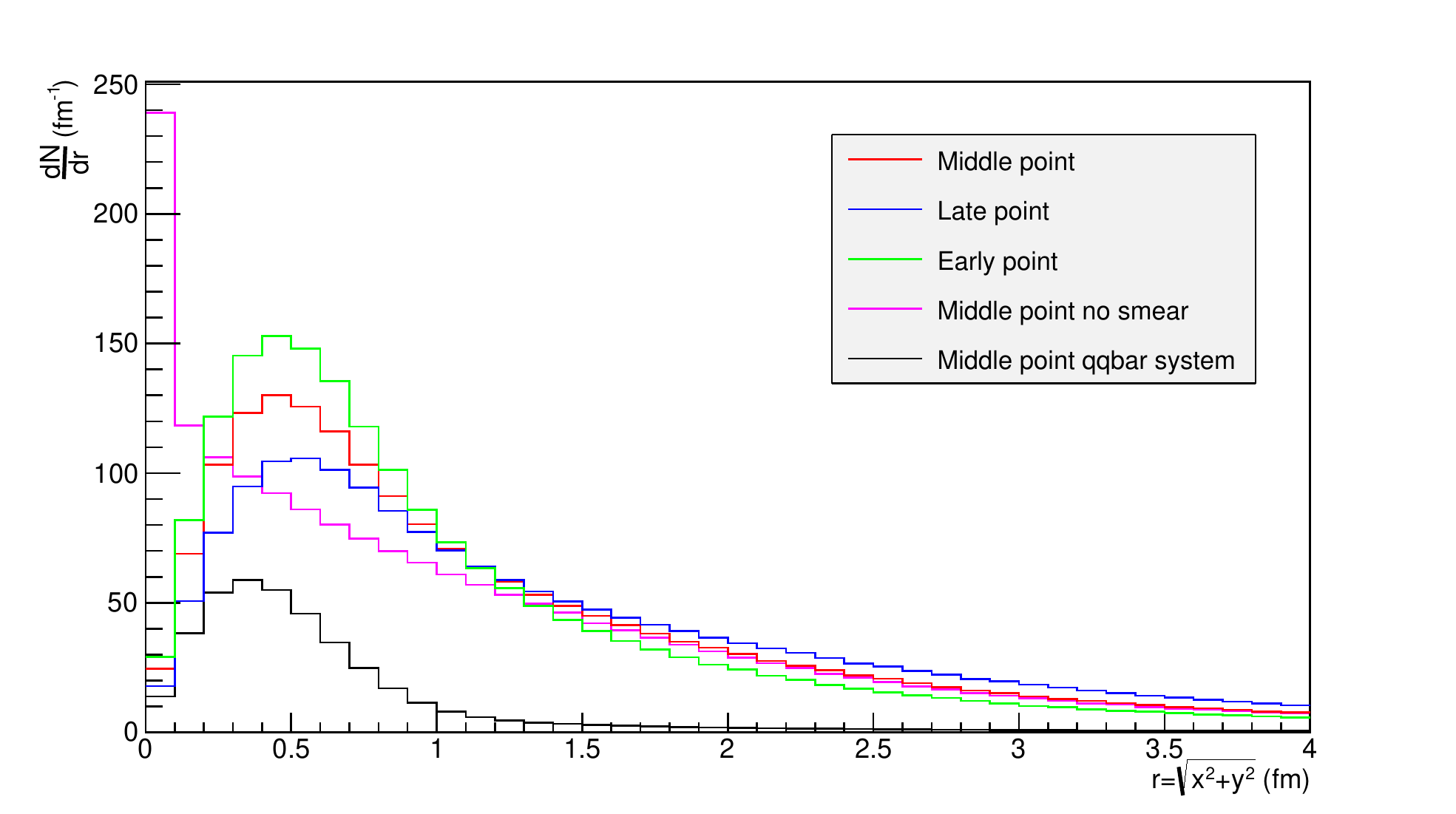}
\captionsetup{justification=centering}
\caption{Transverse spectra for $\p\p$ events and
$\q\qbar$ systems, both at $\sqrt{s}=13$~TeV.}
\label{fig:transverse}
\end{figure}

Figs.~\ref{fig:longitudinal} and \ref{fig:transverse} display the
longitudinal and transverse spectra for $\p\p$ collisions at
$\sqrt{s} = 13$~TeV given by the ``early'', ``middle'' and ``late''
definitions of hadron production points, represented in green, red
and blue, respectively. In the same figures, the spectra for a single
string, at the same CM energy, using the ``middle'' definition are
also illustrated in black. Both primary and secondary hadrons are
taken into account.

The longitudinal spectra for the different definitions are very
similar, as can be seen in Fig.~\ref{fig:longitudinal}. The largest
disagreement is visible around $y_{\tau} \approx 0$, where the spectra
of the ``early'' definition peaks more, but ``early'' also has more
particles at the very largest $y_{\tau}$ values. In short, the
``early'' alternative maximizes the extreme behaviour of hadron
production, whereas the ``late'' one minimizes it. The differences
are not bigger than that we can consider the ``middle'' definition a
fairly reliable one.

Similar conclusions are drawn from the transverse spectra, shown in
Fig.~\ref{fig:transverse}. The  spectrum for $\q\qbar$ events is a
consequence of the transverse smearing, section~\ref{sec:smearing},
and of particle decays; otherwise primary production would all be at
$r=0$. In contrast, $\p\p$ events are constructed out of a large number
of strings stretched between the partons from hard collisions, parton
showers and beam remnants, all of them intrinsically with a transverse
motion. Therefore the smearing is important for the spectrum at low
$r$ values, as can be seen in the difference between the two ``middle''
$r$ distributions, while the distribution at larger $r$ values is
rather insensitive. The difference between the ``early'', ``middle''
and ``late'' production points is larger than for the longitudinal
spectra, but still sufficiently close as to give confidence that
meaningful results can be obtained. In the following, all plots will
be for the ``middle'' definition.

\subsection{Temporal and radial evolution of hadron production}
\label{sec:timeevolution}

\begin{figure}[tp]
\centering
\includegraphics[scale=0.68]{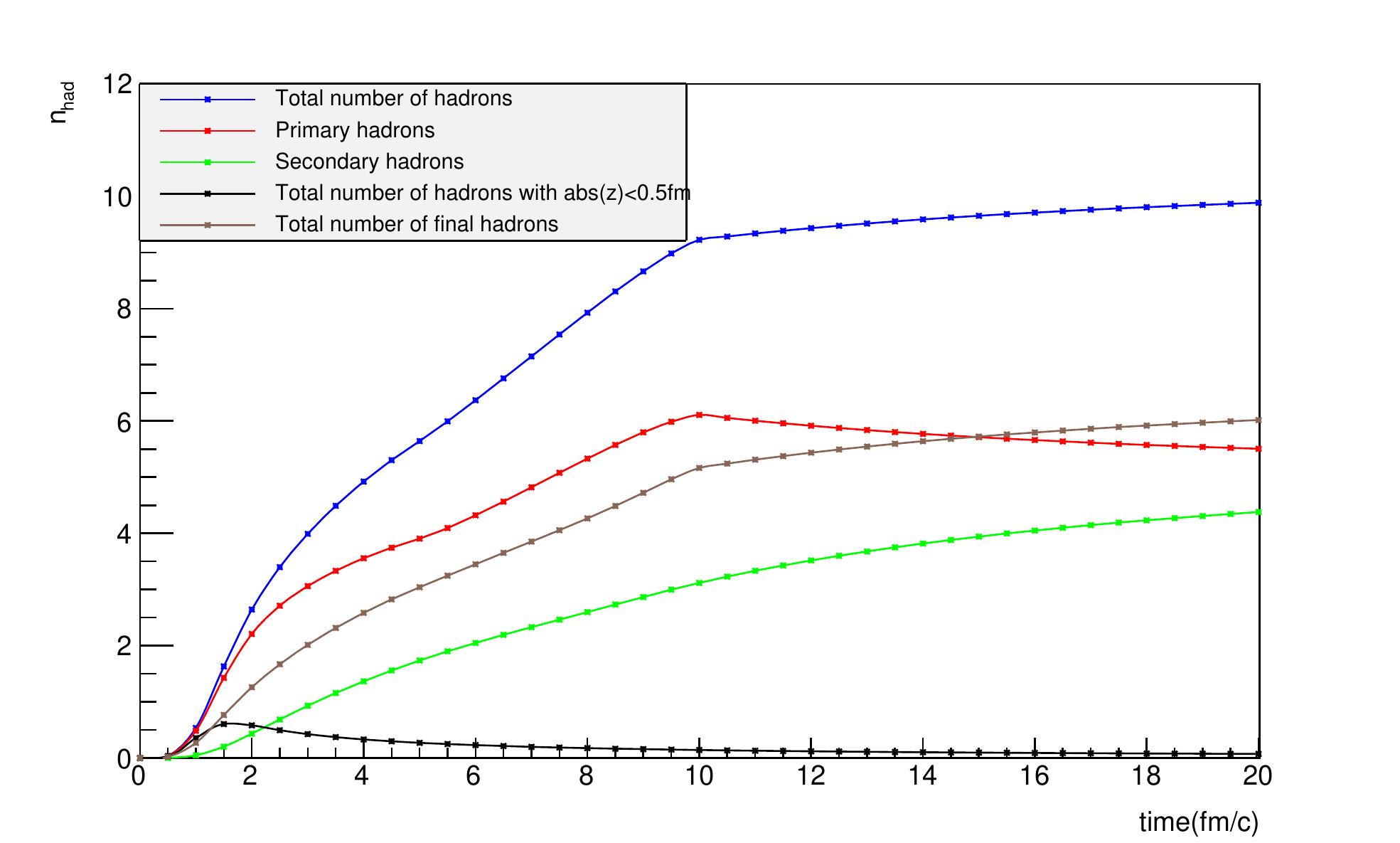}
\caption{Hadron number per event as a function of time for a simple
$\q\qbar$ system formed by massless quarks in the CM frame
with $\sqrt{s}=20$~GeV.}
\label{fig:hadronspertimeqqbar}
\end{figure}

The number of hadrons is shown as a function of time for a single
string with $\sqrt{s} = 20$~GeV in Fig.~\ref{fig:hadronspertimeqqbar}.
The red curve corresponds to the number of primary hadrons, formed
by the string fragmentation, that have not decayed at the time, while
the green curve represents the number of secondary hadrons, from particle
decays. The total number of hadrons, illustrated in blue, is the sum of
primary and secondary hadrons. The brown curve represents the number of
final (i.e.\ stable) hadrons, see section~\ref{sec:junctionsetc}.
Finally, the black curve depicts the number of hadrons with
$|z| < 0.5$~fm, to be discussed in
section~\ref{sec:closepackingproduction}.

For the $20$~GeV simple $\q\qbar$ system in its rest frame, the string
can at most extend 10~fm in the $\pm z$ direction (for $\kappa = 1$~GeV/fm).
This happens at $t = 10$~fm, since the massless quarks move with the
speed of light. The primary hadron production therefore must stop at
this time, as visible in Fig.~\ref{fig:hadronspertimeqqbar}.
Decays make the number of hadrons continue to rise also beyond this
time, but only slowly. Actually many hadrons, like the $\rho^{\pm,0}$
ones, are so short-lived that they decay within some fm of having been
produced.

\begin{figure}[tp]
\centering
\captionsetup{justification=centering}
\includegraphics[scale=0.68]{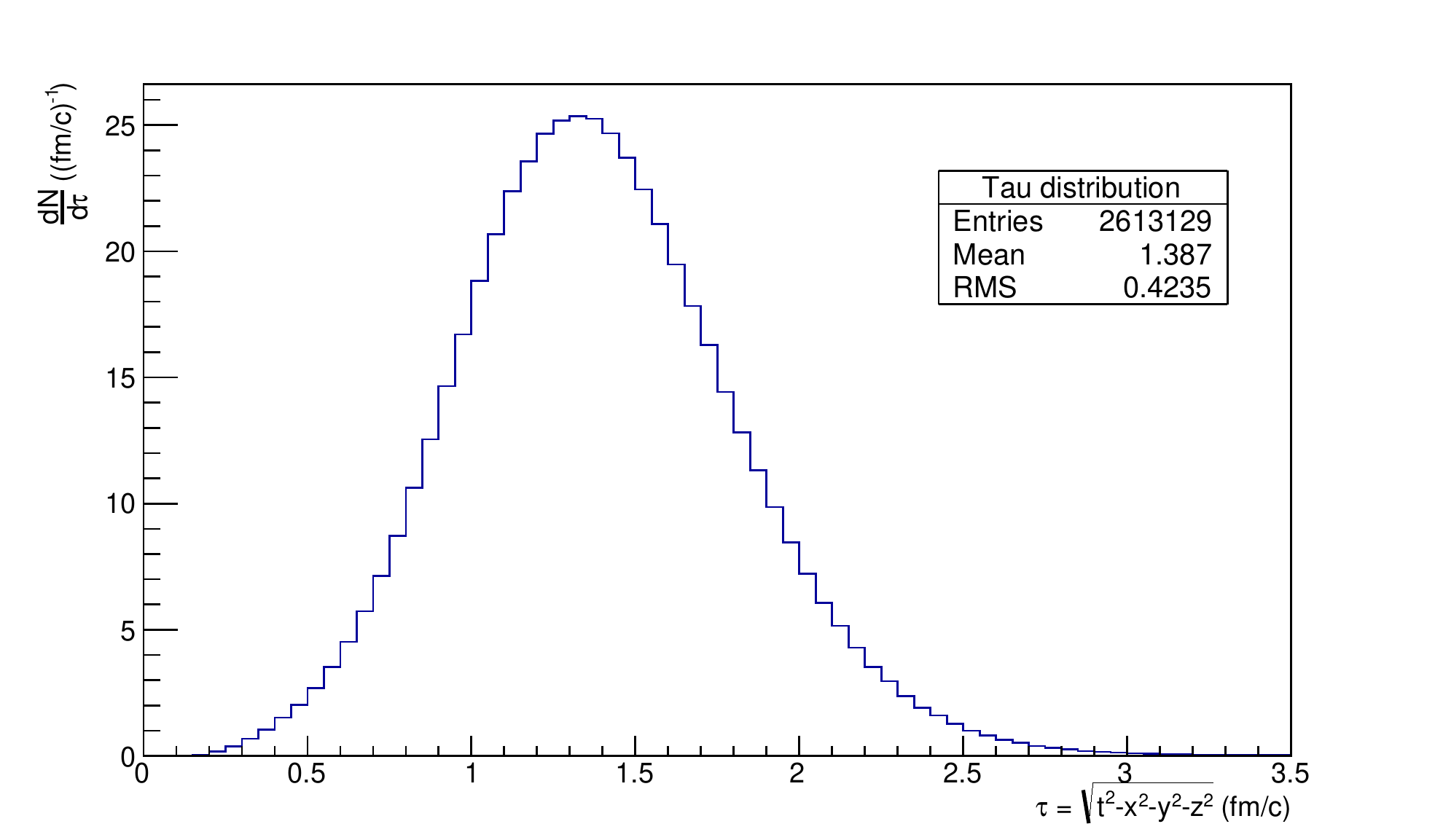}
\caption{Invariant time $\tau$ distribution of primary hadrons
in $\q\qbar$ systems.}
\label{fig:primarytau}
\end{figure}

Note that there are almost no hadrons in the system up until
$t\approx 0.5$~fm, since the string has to have time to begin
stretching out before it can begin to
fragment. This is further illustrated in Fig.~\ref{fig:primarytau},
with the invariant time distribution of primary hadron production
points in the $\q\qbar$ system. By default, the parameters $a$ and $b$
in eqs.~(\ref{eq:probabilityofz}) and (\ref{eq:probabilityofgamma})
are set to $a=0.68$ and $b=0.98$~GeV$^{-2}$ \cite{Skands:2014pea},
giving rise to a suppression of small $\Gamma$ values of breakup
vertices, and thereby also of small hadron production times. In detail,
the relation between $\Gamma$ and $\tau$, eq.~(\ref{eq:gamma}), implies
$P(\Gamma) \propto \Gamma^a \d\Gamma \propto \tau^{2a} \, \tau \, \d\tau %
= \tau^{2a+1} \, \d\tau$ for $\tau \rightarrow 0$.
Furthermore, the expectation value of
$\langle \Gamma \rangle = (1 + a) / b \approx 1.7$~GeV$^2$
gives $\langle \tau \rangle \approx %
\sqrt{ \langle \Gamma \rangle} /\kappa \approx 1.3$~fm, in agreement
with Fig.~\ref{fig:primarytau}. Because those aspects are typical of
the fragmentation process, a similar behaviour is also observed in
$\p\p$ collisions.

\begin{figure}[tp]
\centering
\includegraphics[scale=0.75]{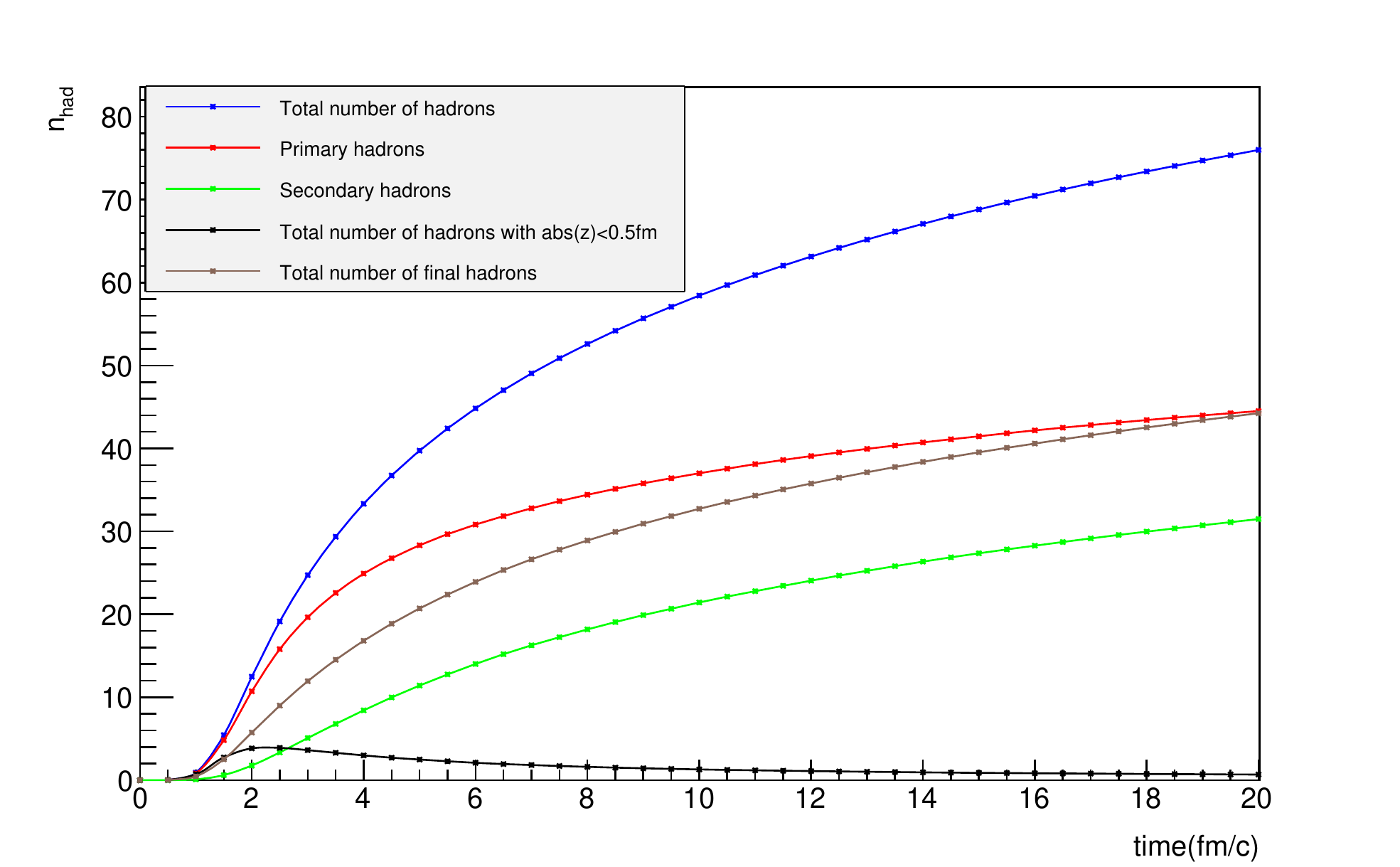}
\captionsetup{justification=centering}
\caption{Hadron number per event as a function of time, up until
$t = 20$~fm, for $\p\p$ collisions at $\sqrt{s}=13$~TeV.}
\label{fig:hadronspertimepp20}
\end{figure}

The time evolution of hadron production in 13~TeV $\p\p$ events is
shown in Fig.~\ref{fig:hadronspertimepp20} for $t \leq 20$~fm.
Although the qualitative behaviour is similar to the one
in $\q\qbar$ systems, the temporal evolution is smoother and the number
of hadrons generated per unit time increases more rapidly in the
$\p\p$ case. These effects are direct consequences of the presence of
several string systems in $\p\p$ events, possibly extending all the way
out to 6500~fm from the origin.

\begin{figure}[tp]
\centering
\includegraphics[scale=0.7]{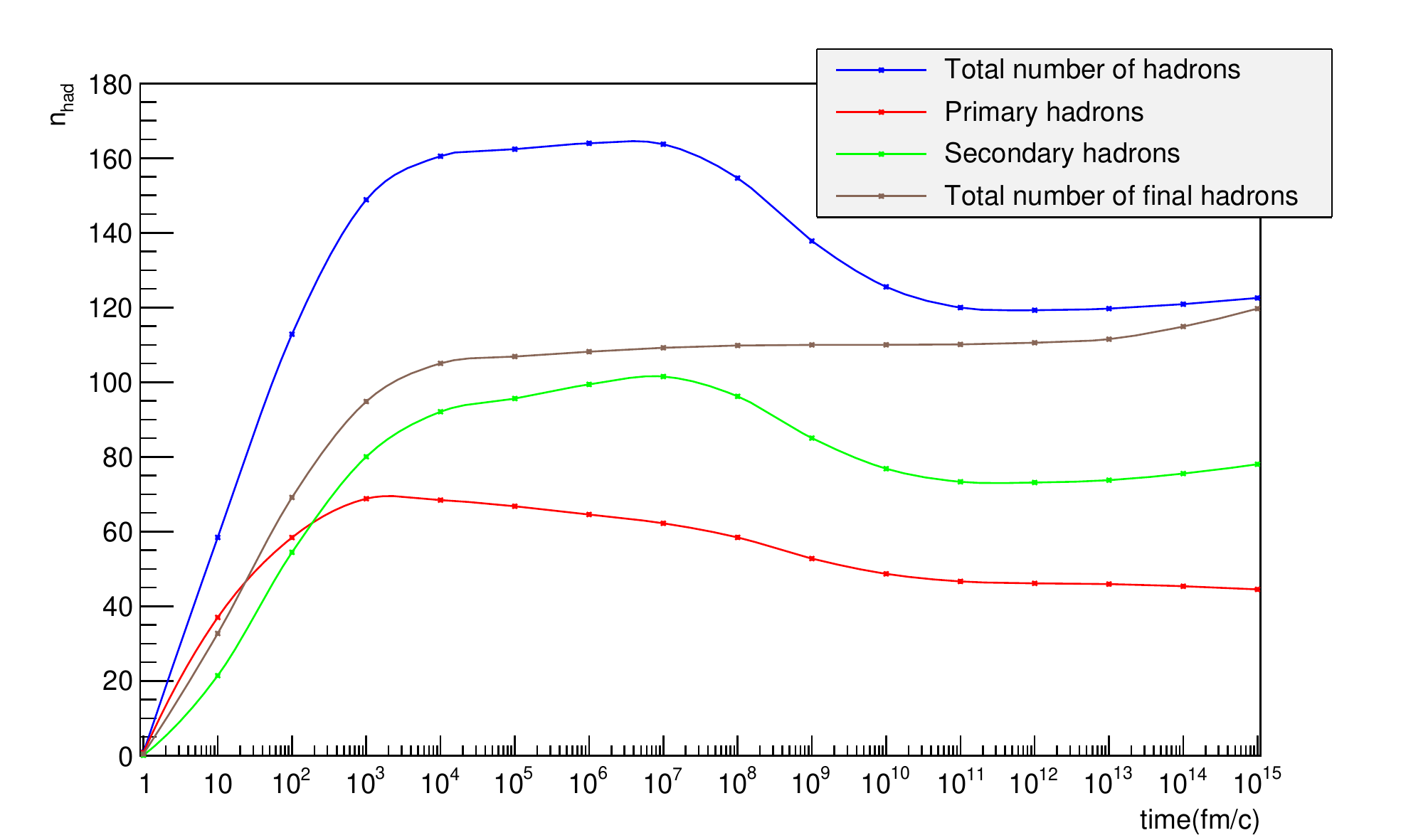}
\captionsetup{justification=centering}
\caption{Hadron number per event as a function of time for 13~TeV
$\p\p$ collisions.}
\label{fig:hadronspertimepp}
\end{figure}

Fig.~\ref{fig:hadronspertimepp} extends the $\p\p$ description up to
$10^{15}$~fm~$= 1$~m. As in the case of the $\q\qbar$ system, the
total number of primary hadrons increases until fragmentation is over,
which now is at $t\approx 10^3$~fm owing to the higher energy.
Decays deplete the number of remaining primary hadrons but increase
the number of secondary ones. The significant drop in the number of
hadrons at $t\approx10^8$~fm is from electromagnetic decays of the
$\pi^0$, mainly $\pi^0 \to \gamma \gamma$. Although the lifetimes of
$\s$, $\c$ and $\b$ hadrons typically are at the mm to cm scales
(more long-lived ones, like $\K^{\pm}$, being considered stable here),
their decays are still ongoing at 1~m, owing to time dilation of the
frequently fast-moving hadrons.

\begin{figure}[tp]  \centering
\includegraphics[scale=0.75]{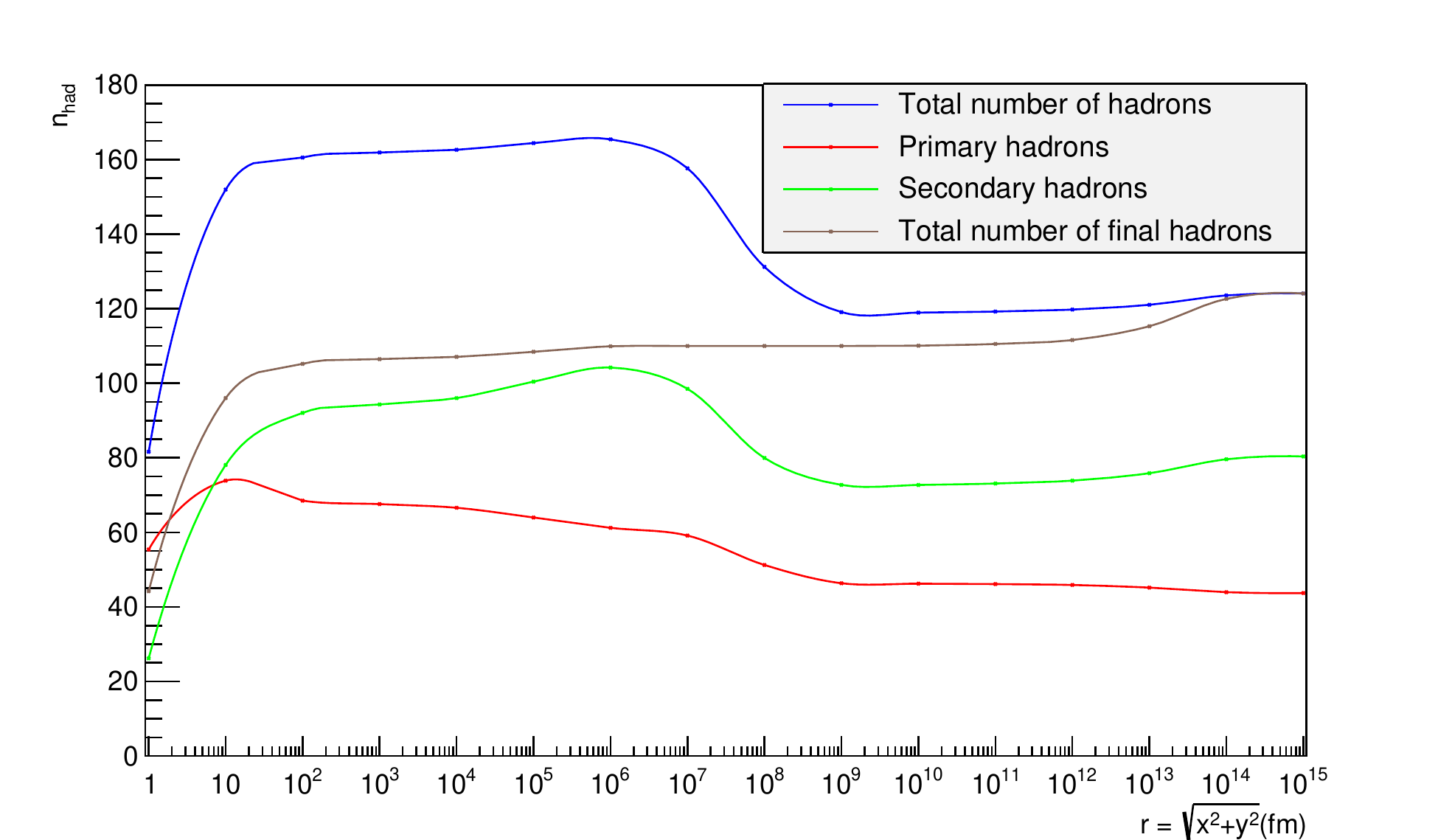}
\captionsetup{justification=centering}
\caption{Hadron number per event as a function or radius for 13~TeV
$\p\p$ collisions}
\label{fig:hadronsperradiuspp}
\end{figure}

Most of the expansion of the system is along the $z$ axis,
i.e.\ the $|z|$ distribution of hadron production would look
similar to the $t$ one in Fig.~\ref{fig:hadronspertimepp},
except for the lack of a suppression at $z = 0$.
It is therefore interesting to show the radial evolution separately,
Fig.~\ref{fig:hadronsperradiuspp}, for the same $t$ range.
Overall the two figures resemble each other, but all the relevant
features have been compressed owing to the lower radial velocities.
The $\pi^0 \to \gamma \gamma$ decay is shifted from $t\approx10^8$~fm
to $r \approx 10^6$~fm, for instance. The impact of weak $\s$, $\c$ and
$\b$ hadron decays are better visible in the range between 1 and 100~mm;
beyond that scale essentially all relevant decays have already
occurred. At the other end of the scale, note that around half of
the hadron production occurs in $r < 1$~fm; there is no equivalent
dynamical suppression of small $r$ as there is of small $t$.

\subsection{Close-packing of hadron production in the central region}
\label{sec:closepackingproduction}

\begin{figure}[tp]
\centering
\subfloat[20 GeV single string systems]{
    \includegraphics[scale=0.75]{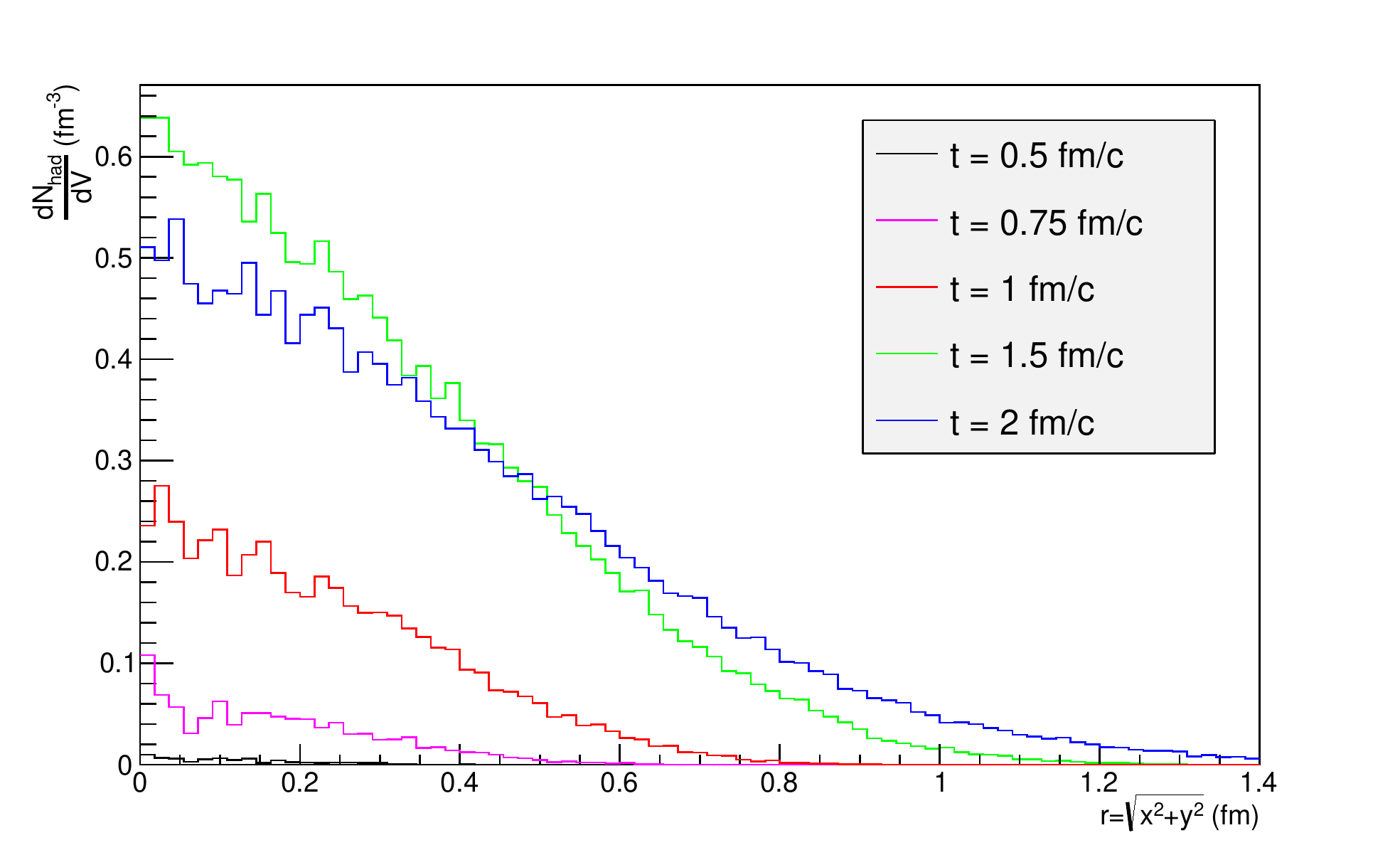}}
\hspace{5mm}
\subfloat[13~TeV $\p\p$ collisions]{
    \includegraphics[scale=0.74]{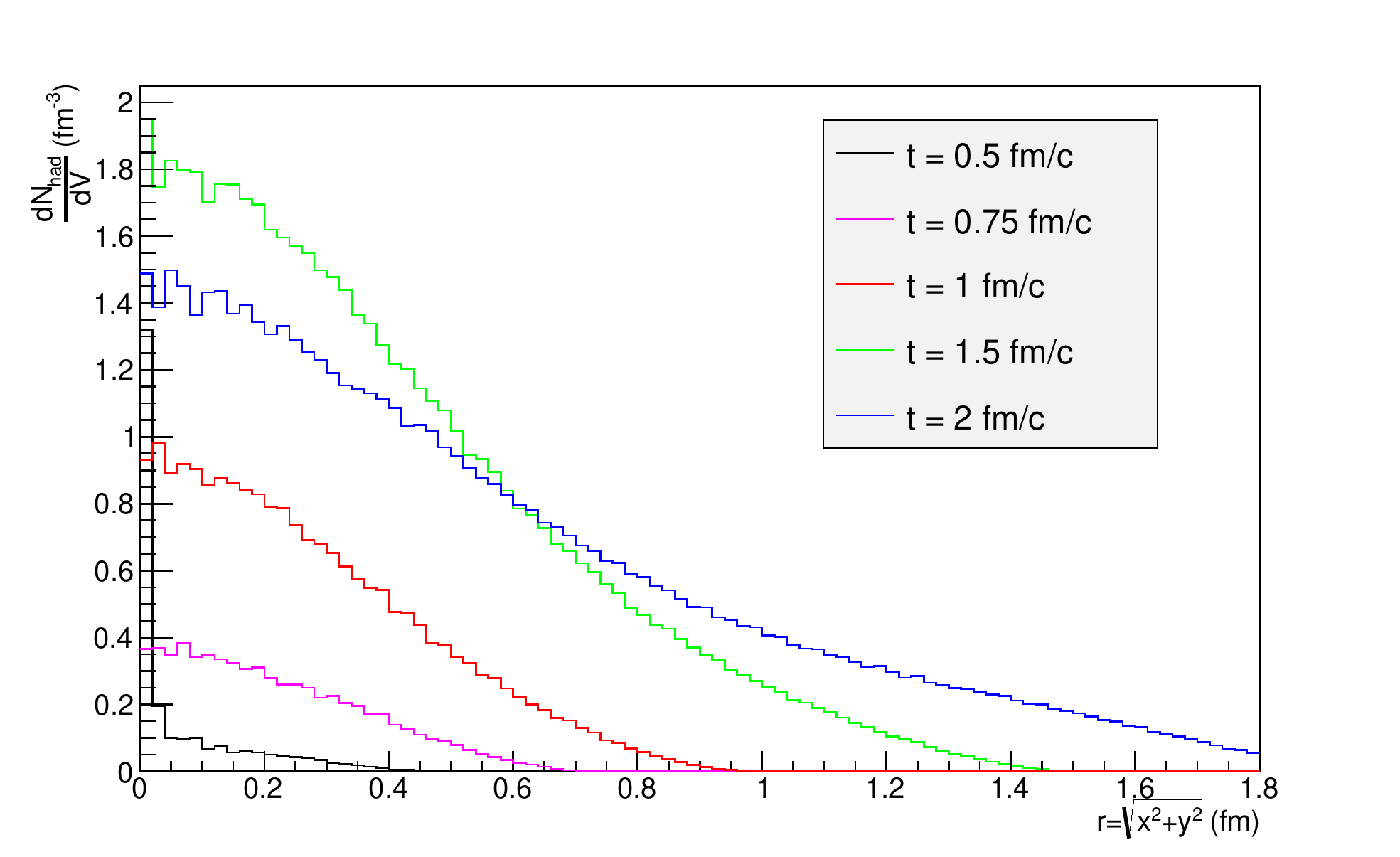}}
\caption{Hadronic density as a function of the radius for different
constant times, for a central slice $|z| < 0.5$~fm.}
\label{fig:densityconstanttime}
\end{figure}

One of the key objectives of this article is to assess the space--time
density of hadron production, $\d N / \d V$. Eventually we will need to
use Lorentz invariant quantities, but these will then hide the time
aspect of the evolution. To begin with, we will therefore study the
density for $|z| \leq 0.5$~fm as a function of $r$ and $t$,
\begin{equation}
\left. \frac{\d N}{\d V} \right|_{|z| \leq 0.5}
= \left.\frac{\d N}{\d x \, \d y \, \d z} \right|_{|z| \leq 0.5}
= \frac{\d N}{\d x \, \d y} = \frac{\d N}{2\pi \, r \, \d r} ~,
\label{eq:densityz05}
\end{equation}
giving a measure of the hadronic densities as a function of radius.
The $r$-integrated number as a function of $t$ is shown in
Figs.~\ref{fig:hadronspertimeqqbar} and \ref{fig:hadronspertimepp20}.
This number only increases up to $t \approx 2$~fm, a time after which
the longitudinal expansion leads to a steady decrease. Therefore,
in Fig.~\ref{fig:densityconstanttime}a, the $r$ distribution is only
shown for a few different $t \leq 2$~fm. The hadron density at times
$t = 0.5$~fm is extremely low both for 20~GeV $\q\qbar$ systems and for
13~TeV $\p\p$ events, since they hardly have had time to start
hadronizing yet. From this point on, hadrons are generated from
fragmentation and particle decays, giving an increasing hadron density
in the central region. The maximal value is at $t \approx 1.5$~fm,
a value that relates well with  typical hadronization time scales,
and where the density at $r = 0$ approaches 2 hadrons per fm$^3$.
A proton has a volume $V_h = 4\pi r_{\p}^3 / 3 \approx 2.76$~fm$^3$
if we use $r_{\p}=0.87$~fm \cite{Patrignani:2016xqp} so, assuming the
same volume for all hadrons and disregarding potential Lorentz contraction
effects, this implies that five hadrons overlap in the center of the
collisions. That number increases rather slowly with the collision energy;
it is around four hadrons at 2~TeV and seven at 100~TeV. Also other
measures of close-packing are expected to display only a mild energy
dependence, so our results at 13~TeV should offer guidance for 
a wide range of collider energies.

\subsection{Hadron production at different multiplicities}
\label{sec:hadronproductionmultiplicities}

In order to extend the previous analysis to a Lorentz invariant measure
of hadronic density, instead the volume element $\d^{3}x / t$ will now be
used:
\begin{equation}
t \, \frac{\d N}{\d^3 x} = \frac{\d N}{\d^{2}r \frac{\d z}{t}} =
\frac{\d N}{\pi \, \d r^2 \, \d y_{\tau}} \rightarrow
\frac{N}{\pi \, r_m^2 \, \Delta y_{\tau}} ~.
\label{eq:multiplicitystudies}
\end{equation}
In the last step $r_m$ is introduced as the median radius of the hadron
creation vertices in the event and $\Delta y_{\tau}$ is the full width at
half maximum of the $\d N / \d y_{\tau}$ distribution. Together $r_m$ and
$\Delta y_{\tau}$ thus define a characteristic volume over which much
of the production will occur, and relate it to a typical maximum density.
For instance, the $|y_{\tau}|$ distribution is roughly triangular in shape,
cf.\ Fig.~\ref{fig:longitudinal}, so $N / \Delta y_{\tau}$ is about
the height of the $\d N / \d y_{\tau}$ distribution at its maximum.

Note that the hadronic multiplicity studied here is different from
typical experimental definitions, e.g.\ the charged multiplicity
in vertex detectors. Since we are interested in the hadronization
process, only strong decays should be taken into account in our
analysis. This excludes electromagnetic and weak decays, such as the
$\pi^0$ one, but furthermore decays with $r > 10$~fm are not taken
into account, since beyond that hadronic densities have fallen to
modest levels anyway. In order to avoid double-counting of a hadron
and its decay products, all secondary hadronic decay vertices enter
with a weight one less than the hadronic multiplicity of the decay.
Counted this way, the average multiplicity of inelastic nondiffractive
13~TeV $\p\p$ events is $n_{\mathrm{had}} = 169$.

\begin{figure}[tp]
\centering
\includegraphics[scale=0.7]{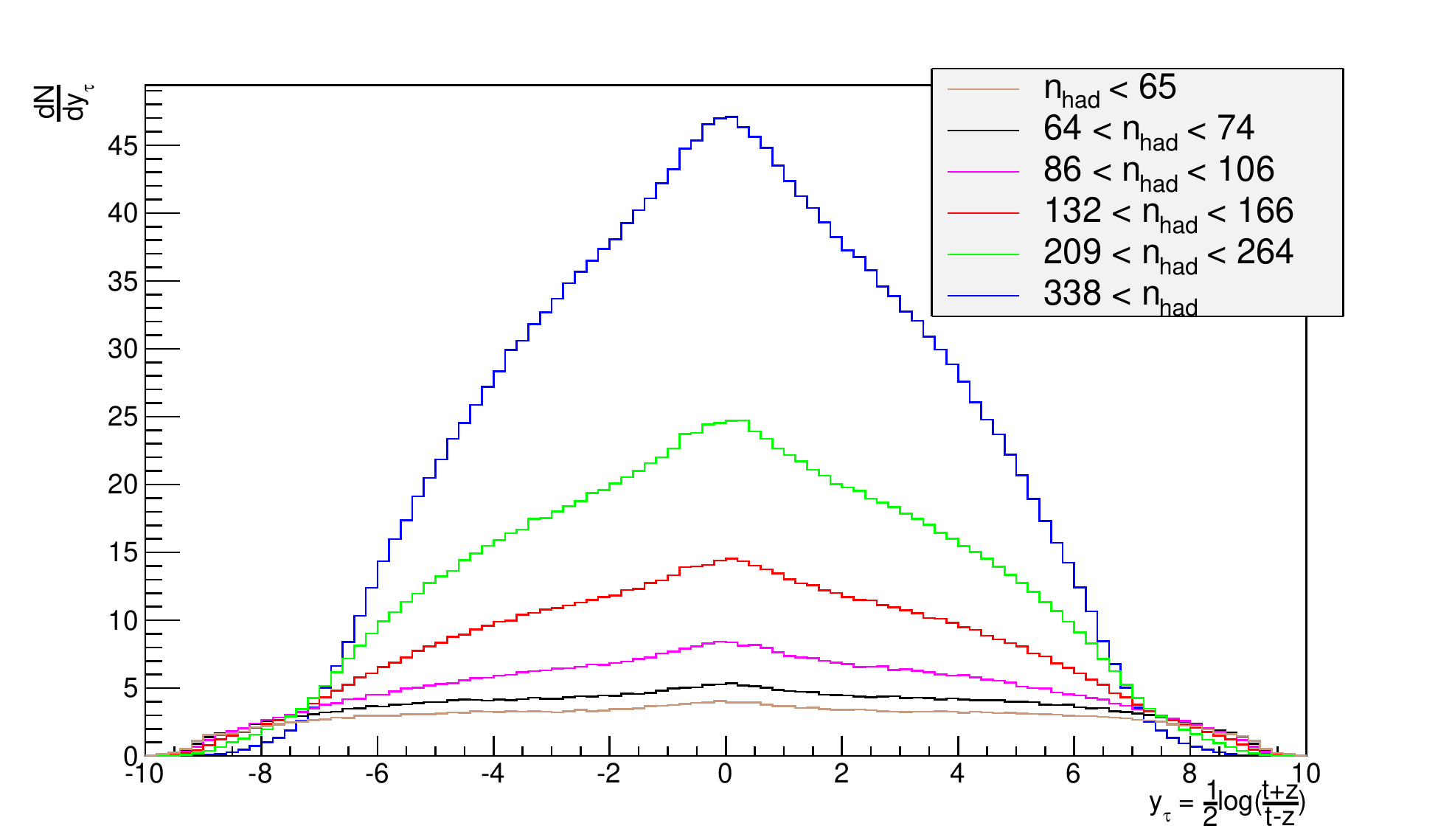}
\captionsetup{justification=centering}
\caption{Longitudinal spectra for 13~TeV $\p\p$ collisions and
different multiplicity ranges.}
\label{fig:longitudinalmult}
\end{figure}

\begin{figure}[tp]
\centering
\includegraphics[scale=0.7]{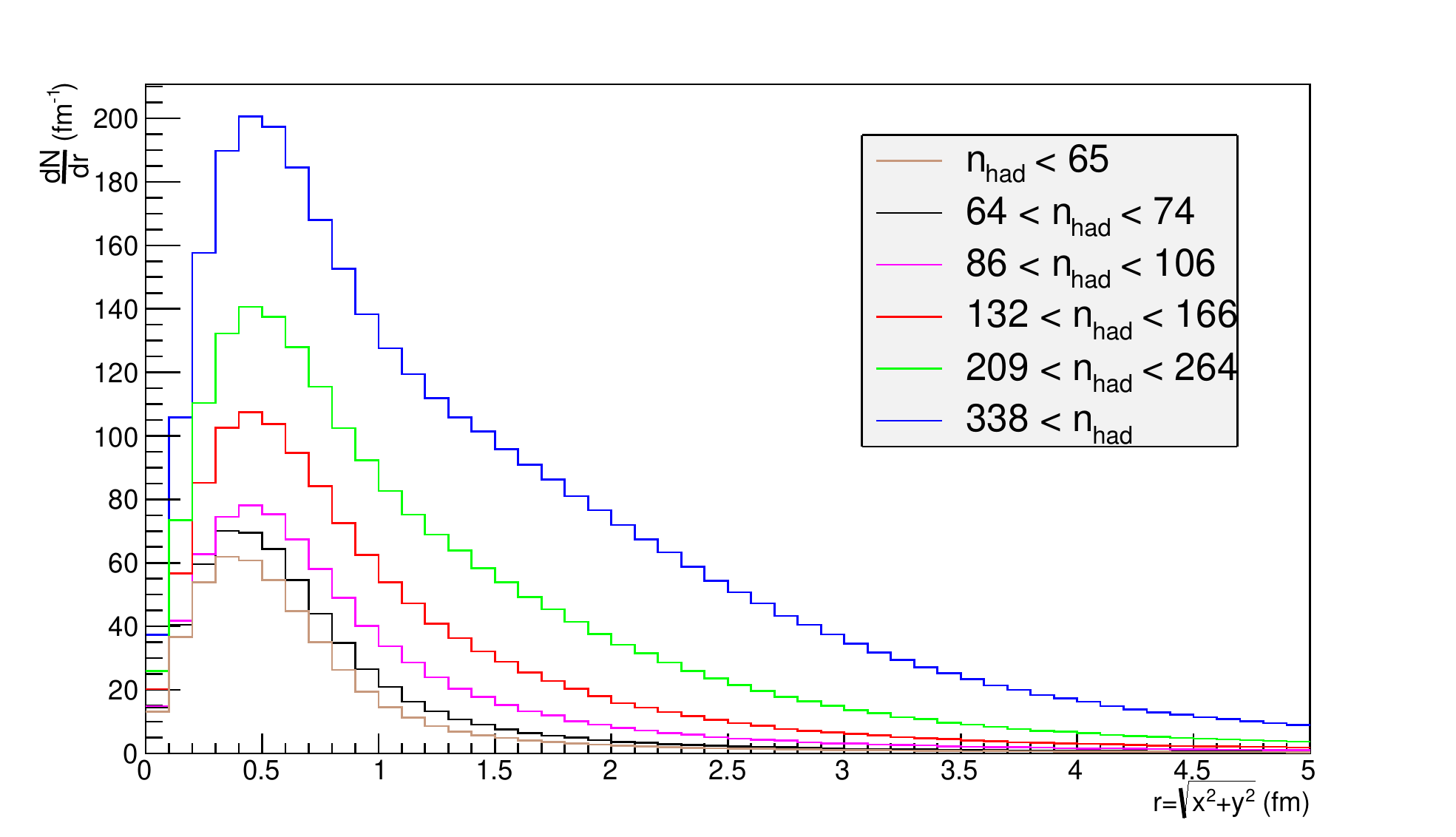}
\captionsetup{justification=centering}
\caption{Transverse spectra for 13~TeV $\p\p$ collisions and different
multiplicity ranges.}
\label{fig:transversemult}
\end{figure}

Inside this sample, ten multiplicity ranges are defined such that each
of them corresponds to roughly 10\% of the events. The resulting
longitudinal $y_{\tau}$ and transverse $r$ spectra are presented in
Figs.~\ref{fig:longitudinalmult} and \ref{fig:transversemult},
respectively. For the sake of clarity, some intermediate multiplicity bins
are left out of the figures. By energy--momentum conservation the $y_{\tau}$
(and $y$) spectra are more peaked around the middle for increasing
multiplicities. Not so for the $r$ spectra, where the distribution
shifts towards larger values for the higher multiplicities.
It is here useful to remind that the basic MPI framework implies that
high multiplicities primarily come from having more MPIs, rather than
e.g.\ from a single hard interaction at a larger $\pT$ scale, and that
therefore $\langle \pT \rangle (n_{\mathrm{charged}})$ is expected to be
reasonably flat. The experimental observation of a rising
$\langle \pT \rangle (n_{\mathrm{charged}})$ actually was the reason
to introduce colour reconnection (CR) as a key part of a realistic
MPI modelling \cite{Sjostrand:1987su}.

\begin{figure}[tp]
\centering
\includegraphics[scale=0.7]{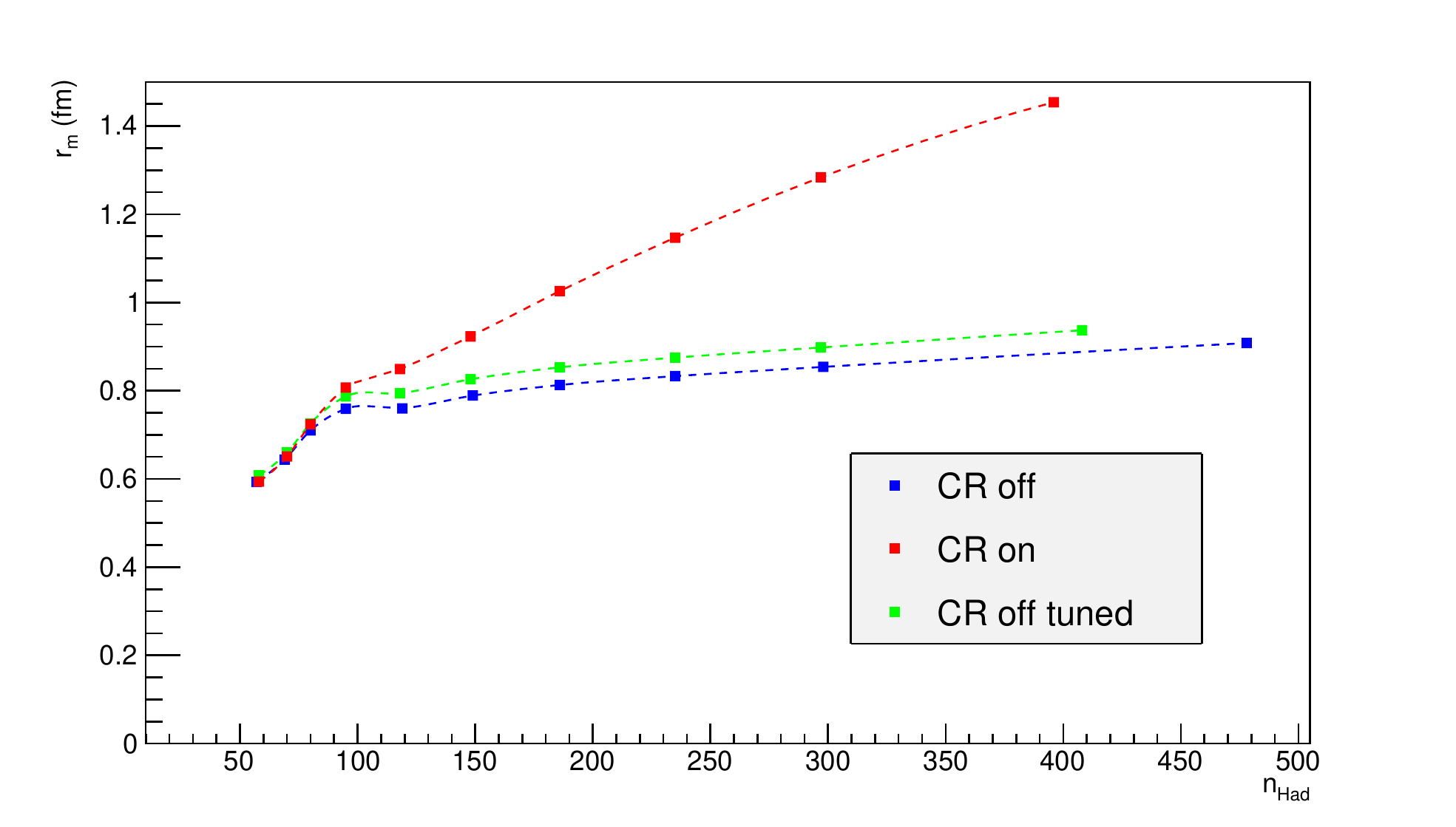}
\caption{Median radii as a function of multiplicity for 13~TeV $\p\p$
collisions. The red curve corresponds to the approach with colour
reconnection, while the blue and green curves represent the model
without colour reconnection and a tuned model without colour
reconnection, respectively.}
\label{fig:radiimultiplicity}
\end{figure}

The effect of CR on the median radii $r_m$ is shown in
Fig.~\ref{fig:radiimultiplicity}, as a function of the median hadronic
multiplicity $n_{\mathrm{had}}$ of each multiplicity range. The red and
blue curves represent results with and without CR, respectively,
and these match very well with expectations from the
$\langle \pT \rangle (n_{\mathrm{charged}})$ behaviour; also the rise
of $r_m$ is driven by the CR mechanism. Note that switching off CR
gives higher event multiplicities, well above data.
To this end also a green curve is introduced, wherein the $\pTo$
parameter of the MPI framework \cite{Sjostrand:2017cdm} is increased
for the no-CR alternative until the average multiplicity is the same
as in the default with-CR scenario. This gives a slightly larger
$r_m$ than the naive no-CR setup, since the $\langle \pT \rangle$
of MPIs is increased in the process, but otherwise is in line with
the original observation.

\begin{figure}[tp]
\centering
\includegraphics[scale=0.7]{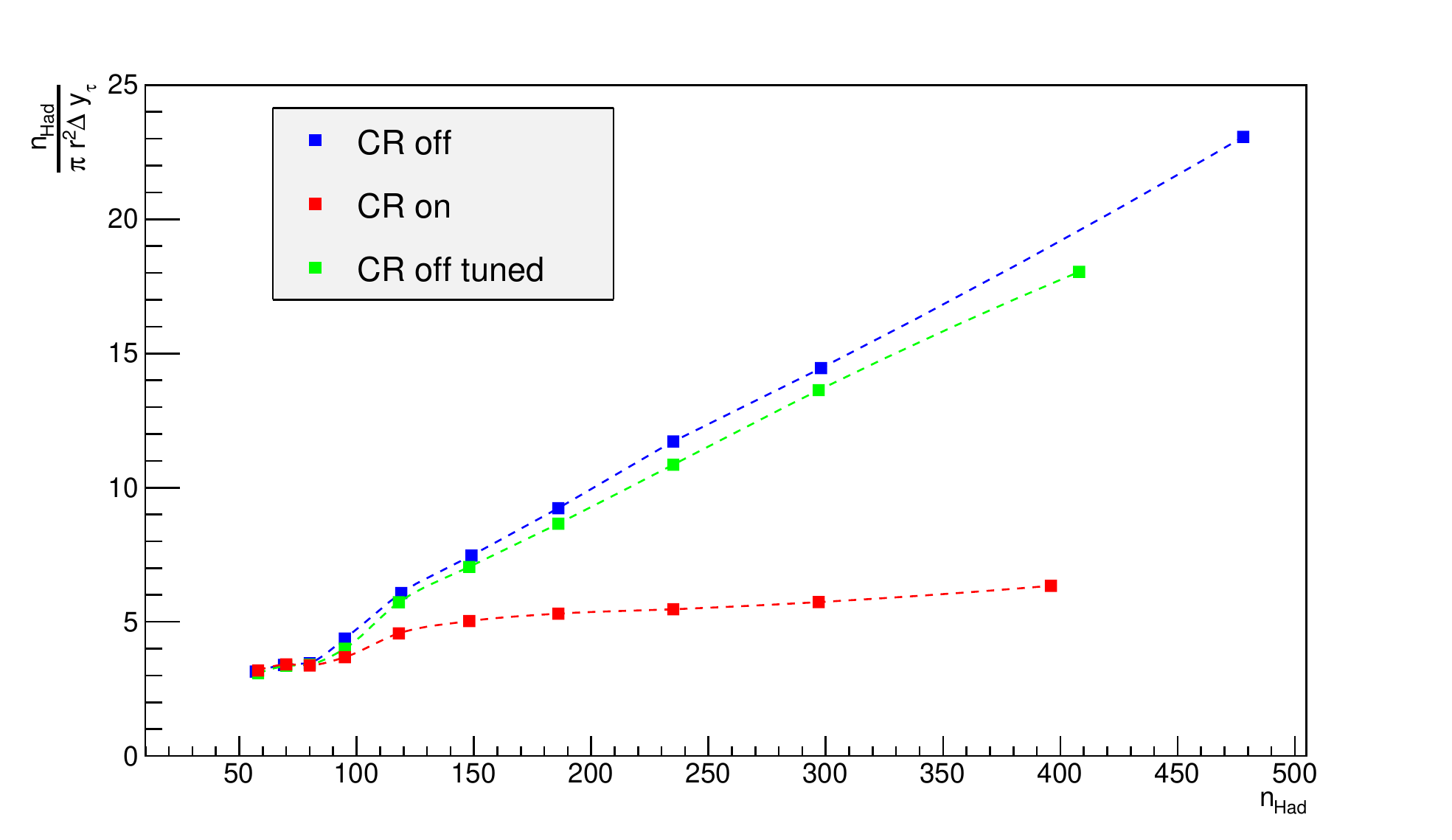}
\caption{Hadron density as a function of multiplicity for $\p\p$
collisions at 13~TeV. The red, blue and green curves represent the
three different models with and without colour reconnection, also
included in Fig.~\ref{fig:radiimultiplicity}.}
\label{fig:hadprodmult}
\end{figure}

Fig.~\ref{fig:hadprodmult} shows the hadron density, defined as in
eq.~(\ref{eq:multiplicitystudies}), for the three same scenarios as
above. The $n_{\mathrm{had}}$, $r_m$ and $\Delta y_\tau$ are
calculated in each multiplicity range. The space--time hadron
density increases with hadronic multiplicity, but significantly faster
in the two scenarios without CR, as a direct consequence of the
inverse quadratic dependence on $r_m$. The lower values with CR on
may be partly misleading, however; only because strings are spread
across a bigger transverse area when CR is on, it does not mean that
there are strings everywhere in that area. The typical average density
of 5 hadrons per Lorentz invariant space--time element should therefore
be viewed as a lower estimate.

\subsection{Close-packing analysis in the hadron rest frame}
\label{sec:closepackingrestframe}

As a final measure of close-packing we will next check how many hadrons
overlap with each of the hadrons of an event, as defined in the rest
frame of the hadron at the time when it is formed. In detail, consider
a hadron $h_1$ generated at time $t_1$, where $t_1$ is defined in the
rest frame of hadron $h_1$. The other hadrons in the system are boosted
to the rest frame of $h_1$, where only the hadrons created at times
$t \leq t_1$ and which have not decayed at $t_1$ are taken into account.
Their location at $t_1$ is calculated from the respective production point
and four-momentum, from which the distance to $h_1$ can be calculated.
If this distance is shorter than $2r_{\p}$, $r_{\p}$ being the proton
radius, the hadrons are considered to overlap, implying that already
the production of $h_1$ could be affected by the presence of these other
hadrons. Note that Lorentz contraction is not taken into account, which
would decrease numbers, but then neither is the possibility of closer
distances at $t > t_1$, which would increase them. The
analysis is done including or excluding the adjacent hadron on each side
along the string of the hadron studied. The reason for the latter scenario
is that any effects of same-string-neighbours already effectively should
have been taken into account in the tuning of the fragmentation process,
e.g.\ in eq.~(\ref{eq:probabilityofz}).

\begin{figure}[tp]
\centering
\subfloat[Including adjacent hadrons]{
    \includegraphics[scale=0.75]{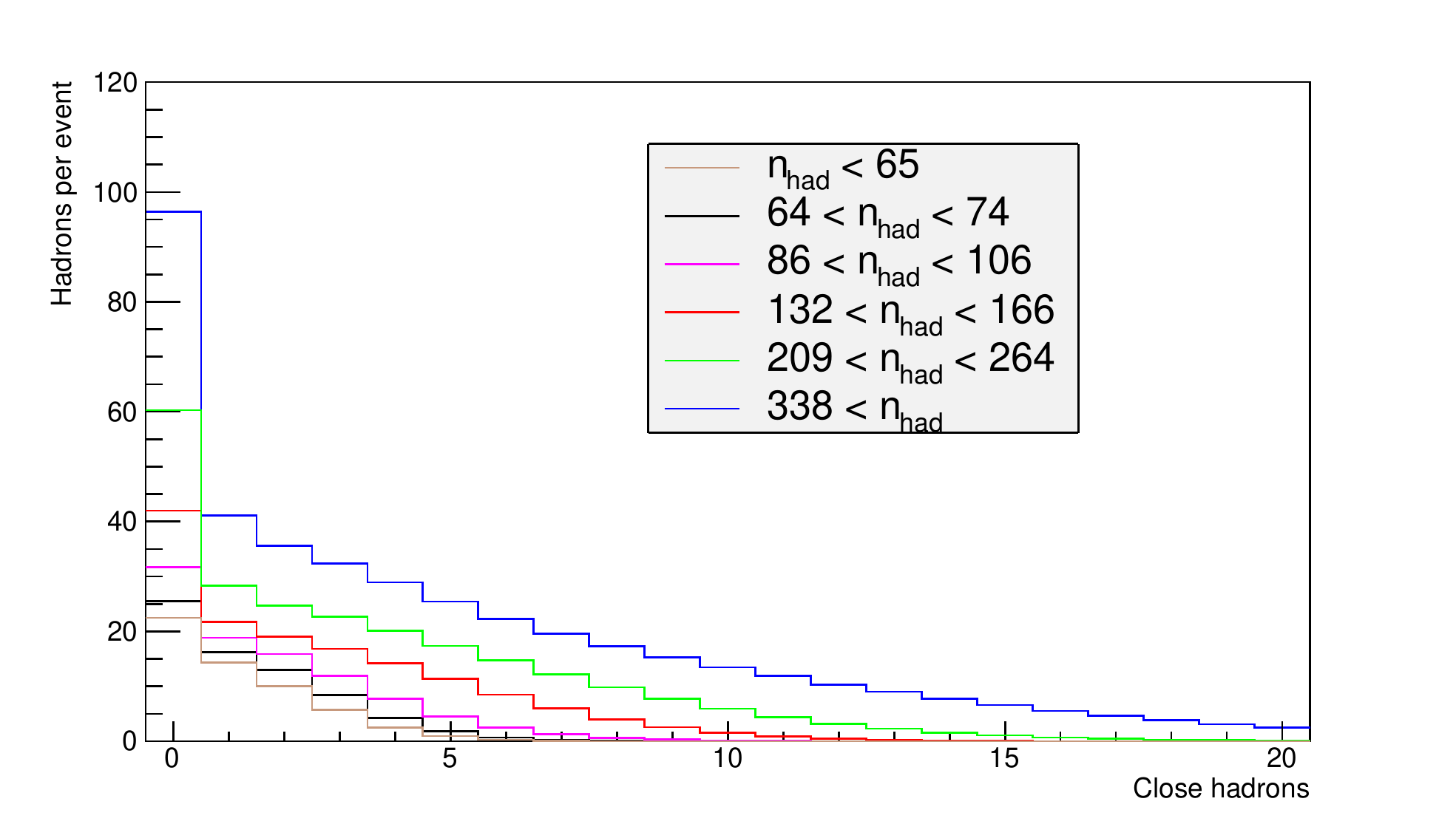}} \hspace{5mm}
\subfloat[Excluding adjacent hadrons]{
    \includegraphics[scale=0.75]{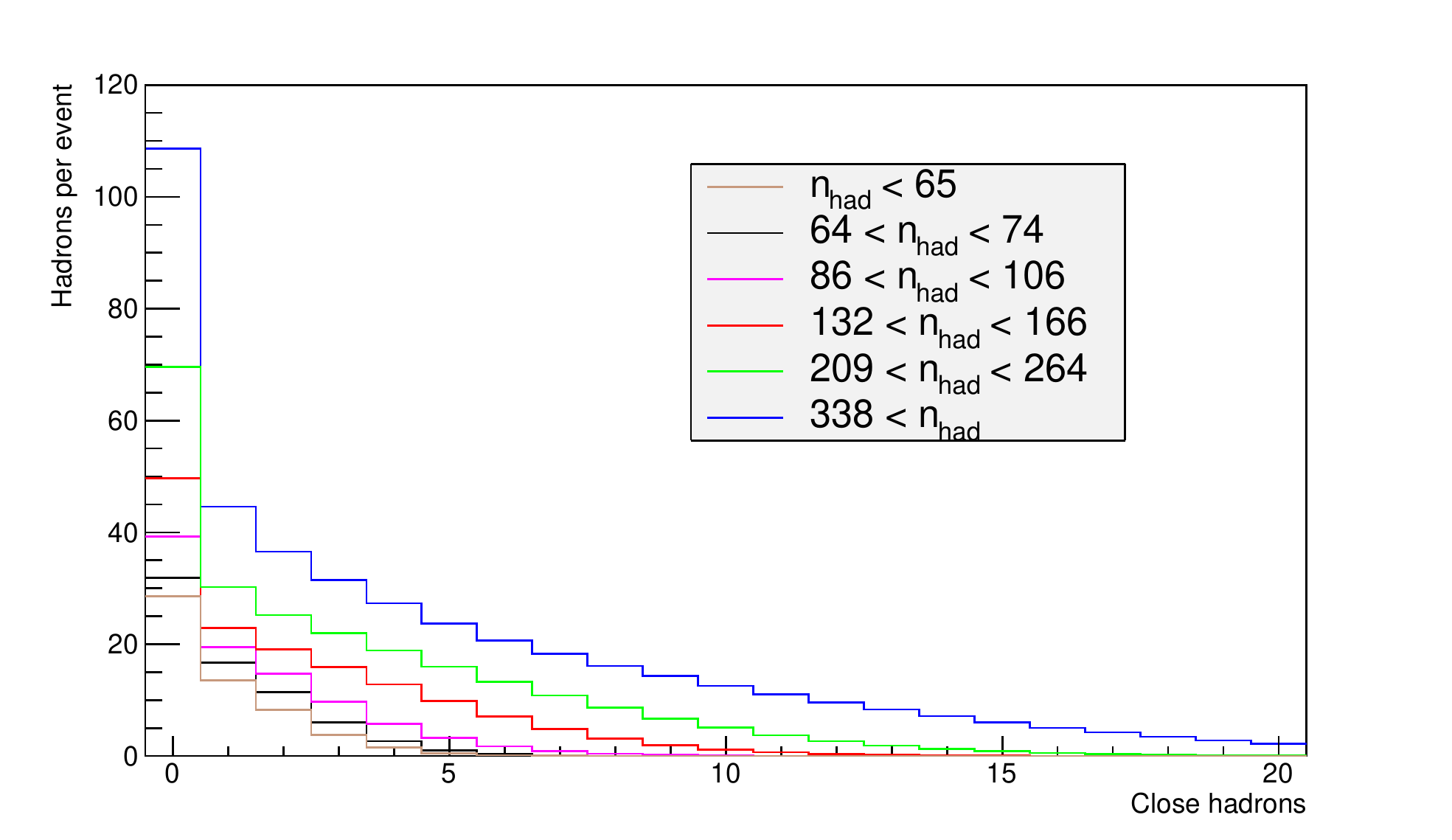}}
\caption{Hadron overlap for different multiplicity ranges for 13~TeV
$\p\p$ collisions.}
\label{fig:closepacking}
\end{figure}

The number of overlapping hadrons is shown in
Fig.~\ref{fig:closepacking} for different hadronic multiplicity
ranges, as presented in section~\ref{sec:hadronproductionmultiplicities}.
Although close-packing also takes place in low-multiplicity $\p\p$ events,
the number of hadrons overlapping with a newly created one is not so high.
For high-multiplicity events, on the other hand, close-packing often arises
with a significant number of nearby hadrons, likely leading to collective
effects that are not taken into account in \textsc{Pythia}.

\begin{figure}[tp]
\centering
\includegraphics[scale=0.8]{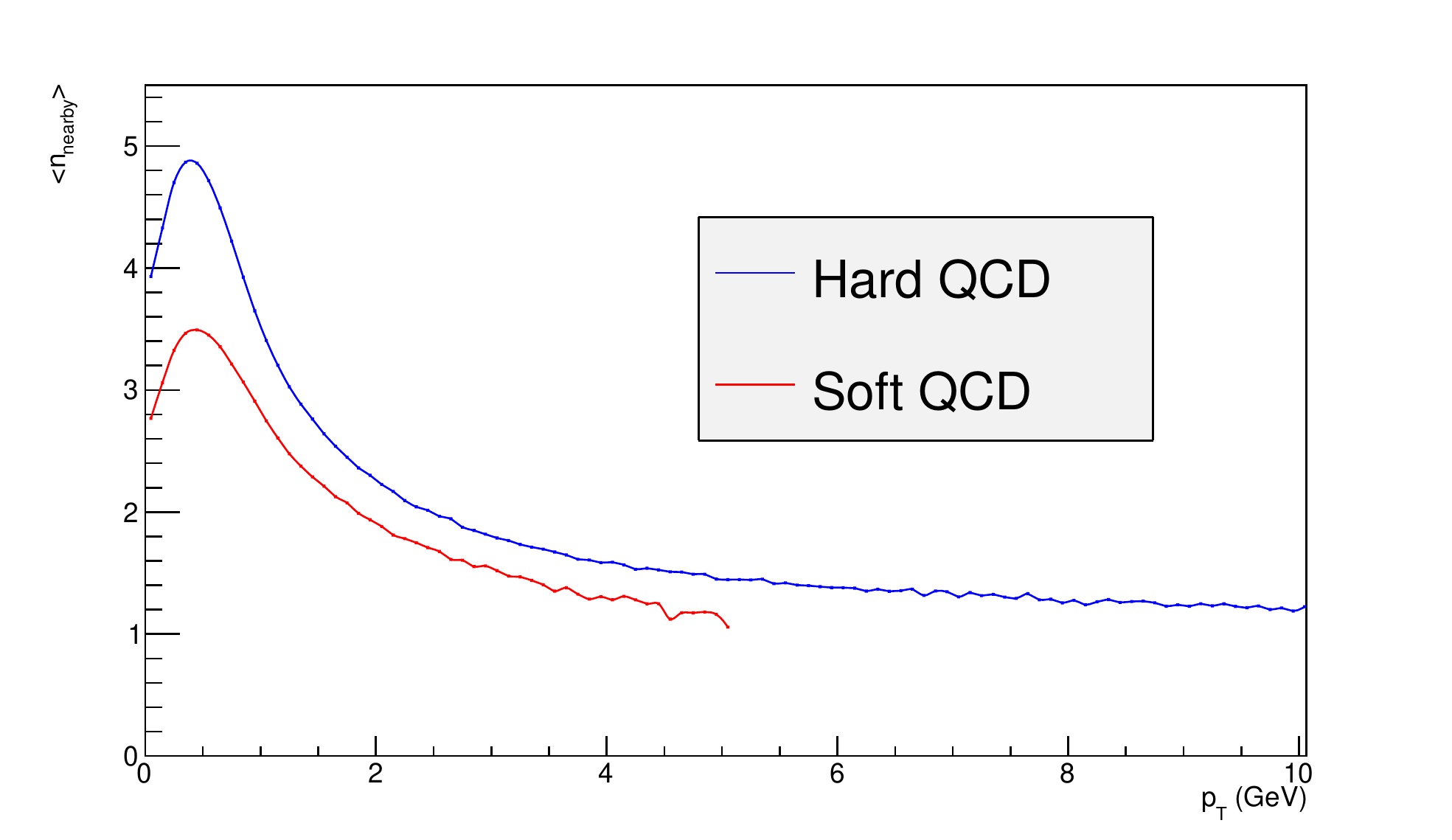}
\caption{Average number of overlapping hadrons as a function of the
$\pT$ of the hadron studied. The red and blue distributions illustrate
the soft and hard QCD processes, where the former stops at 5~GeV
owing to limited statistics at large $\pT$.}
\label{fig:ptsofthard}
\end{figure}

The overlap can be differentiated further. Generally, particles produced
at large transverse momenta are not expected to experience close-packing
as much as those at small ones. The reason is that, even if parton
showers can generate many partons from each initial high-$\pT$
parton, these daughter partons are spread widely in momentum
space. Therefore, the fragmenting strings stretched between them also
will have a modest overlap, unlike the accretion of low-$\pT$
strings from multiple soft MPIs. In order to isolate this feature, we
study the overlap as a function of the hadron transverse momentum,
using the same analysis procedure as above, with exclusion of adjacent
hadrons along the string, Fig.~\ref{fig:ptsofthard}, for ``soft''
and ``hard'' QCD events in red and blue, respectively. The former is
the standard inelastic nondiffractive event sample, whereas the latter
is for the subsample where a hard $2 \to 2$ QCD process has
$\pT > 100$~GeV. In both cases the overlap peaks for hadrons around
$\pT \approx 0.5$~GeV, and then falls off at larger $\pT$ values.
The level is somewhat higher for the hard-QCD events, consistent with
such events being biased towards smaller impact parameters and therefore
more MPIs, but the trends are consistent.

\section{Summary and Outlook}
\label{sec:summary}

The motivation for this article is the mounting evidence for several
collective effects in high-multiplicity $\p\p$ collisions, similar to
those usually associated with the formation of a Quark--Gluon Plasma in
heavy-ion collisions. Whether we are witnessing QGP also in $\p\p$
or not remains an open question, but the need to allow for some kind of
collective mechanisms can hardly be in doubt. It should not even come
as a surprise, given that already order-of-magnitude estimates of
the size of the fragmentation region told us that strings would be
formed close-packed and fragment into close-packed hadrons within any
realistic MPI-based scenario. Colour reconnection was introduced as
a partonic-state mechanism to describe some signals of collectivity,
notably the rise of $\langle \pT \rangle (n_{\mathrm{charged}})$.
But the rising fraction of multistrange baryon production with event
multiplicity implies that collective effects are needed also in or after
the fragmentation stage, or both.

To be able fully to explore various such scenarios it becomes important
to understand the space--time structure of hadronization in more detail
than hitherto. The aim of this article has been to develop the necessary
framework, and implement it as part of the public \textsc{Pythia}
event generator. Specifically, we have determined the space--time location
of the string breakup vertices and compared three alternative definitions
for primary hadron production points. Although the implementation of the
space--time picture in a simple $\q\qbar$ string topology is straightforward,
the picture gets much more intricate when more complicated topologies
are addressed.

To illustrate the usefulness of the new framework, some simple first
studies have been presented, notably exploring space--time hadron densities.
Initially, inclusive longitudinal and transverse space-time distributions
were shown, and the production and decay patterns from fm to m scales
were traced. Next the density in a central slice $|z| < 0.5$ was studied
as a function of $t$ and $r$. While not explicitly Lorentz invariant,
it gave some first hints of close-packing problems. Moving from a volume
element $\d^3 x$ to $\d^3 x / t$ gave access to Lorentz-invariant density
measures. It was shown that the median radius of the fragmentation
region is increasing with multiplicity, but almost only because of the
colour reconnection effects. The flip side is that density is increasing
significantly with multiplicity without CR, whereas it remains at an
average of about five hadrons overlapping with CR included.

The close-packing of hadrons was finally analysed by counting the number
of hadrons overlapping with a newly generated one in its rest frame,
again for different event multiplicities. In this case, the number of
nearby hadrons does increase with multiplicity, with CR included,
implying that close-packing becomes increasingly important with
multiplicity also here. The overlap is largest for low-$\pT$ hadrons, in the
MPI-dominated region, whereas it drops for larger $\pT$ scales,
dominated by hard QCD jets.

A few corners have been cut in the current $\p\p$ implementation. Notably
no space--time vertices have been assigned to the individual MPI
collisions, although such assignments are implicit in the MPI
impact-parameter and matter-profile framework \cite{Sjostrand:2017cdm}.
A sensible space--time picture of parton-shower evolution would
introduce offsets, although presumably not major ones. Similarly, the
CR between different MPIs implies that the two ends of a string may
start out from different space--time points. For now, all such effects
have implicitly been made part of the generic smearing step in
section~\ref{sec:smearing}.

To these minor corrections should be added the potentially much larger
dynamical ones that could generate collective effects, be it before,
during or after the string fragmentation stage. The shove and rope
mechanisms are two examples for the first two stages, but the immediate
continuation of the current article would be to study the consequences
of hadronic rescattering in a dense hadronic gas. Models for hadronic
rescattering already exist \cite{Zhang:2017esm}, such as UrQMD
\cite{Bass:1998ca} and SMASH \cite{Weil:2016zrk}, and could possibly be
interfaced. For better control, however, it would be useful to implement
relevant aspects of such a framework as an integrated part of the
\textsc{Pythia}  program.

The longer-term expectation is that continued experimental studies
will provide further information on all kinds of collective phenomena
in LHC $\p\p$ events, and that model building will try to rise to the
challenge. Especially interesting is to figure out which phenomena can be
explained without invoking QGP, and which cannot. This would then
reflect back on the LHC heavy-ion program.

\section*{Acknowledgements}

Work supported in part by the Swedish Research Council, contracts number
621-2013-4287 and 2016-05996, and in part by the MCnetITN3 H2020 Marie
Curie Initial Training Network, grant agreement 722104.
This project has also received funding from the European Research
Council (ERC) under the European Union's Horizon 2020 research
and innovation programme, grant agreement No 668679.

\appendix
\section{Space--time location of the final breakup}
\label{sec:spacetimefinal}

Fragmentation of an open string is modelled by allowing hadron
production from either string end, until the remaining invariant mass
of the system is only sufficient to generate the last two hadrons (see
section~\ref{sec:breakuppoints}). At that point, a final breakup is
generated between the last previous breakup on either side. For the
energy--momentum picture, the final breakup occurs in a fictitious
final region, created from the combination of all the unused parts of
all remaining regions. This region does not have a space--time
correspondence, however; in particular there is no concept of a
space--time offset where this region is created. In this case we
therefore have had to depart from the energy--momentum picture,
to develop an unfortunately rather complex procedure.

\begin{figure}[tp]
\centering
\includegraphics[scale=0.35]{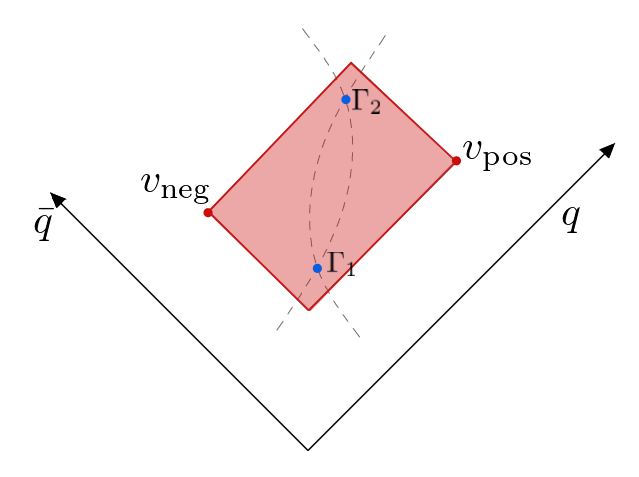}
\caption{The two possibilities for the final breakup, in blue, and the
final region, in red, in a $\q\qbar$ system. The two old
breakups, $v_{\mathrm{neg}}$ and $v_{\mathrm{pos}}$, are also
represented in red.}
\label{fig:finaltwo}
\end{figure}

To set the stage, consider a simple $\q\qbar$ string, where the
flavours of the final breakup helps define the two final hadrons.
The transverse mass constraint of each of those hadrons is represented
by hyperbolae, see section~\ref{sec:howpointsobtained}. The two
hyperbolae either do not cross at all or else cross in two different
points. No solution can be found in the former case, and the
fragmentation process then has to be repeated. The latter case is
illustrated in Fig.~\ref{fig:finaltwo}, where the remaining region is
depicted in red and the blue dots represent the two points
where the hyperbolae meet. Since the two possibilities have different
$\Gamma$ values, the relative probabilities for them to occur are given
by eq.~(\ref{eq:probabilityofgamma}). For simplicity, only the
exponential part is retained, i.e.\ $P(\Gamma_i) \propto \exp(-b\Gamma_i)$
is used to pick either point. That choice made, the kinematics is
fully defined.

Major complications are found in systems with several intermediate
gluons between the $\q$ and $\qbar$ ends, specifically when the two old
breakups are located in different regions. In those cases, knowing the
region in which the final breakup is located is not always possible. Since
that aspect is essential to calculate the $\hx^{\pm}$ fractions, several
methods have been tested before settling on the one presented here.

\begin{figure}[tp]
\centering
\includegraphics[scale=0.35]{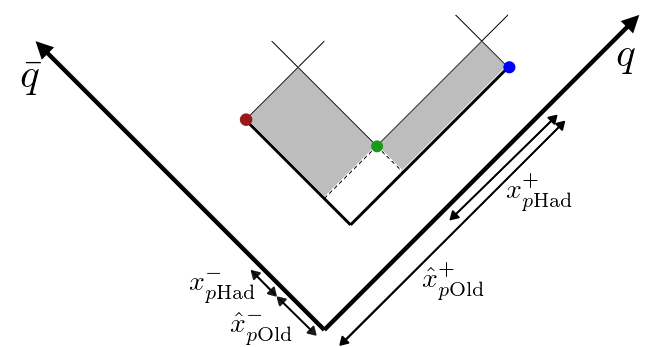}
\caption{Final breakup point and final two hadrons, corresponding to
the grey regions, in a $\q\qbar$ system. The red and blue
points are the previous breakups and the endpoints of the final
region, while the green dots represents the final breakup. }
\label{fig:finaltwoproj}
\end{figure}

For notational convenience, the old breakup closer to the $\q$ ($\qbar$)
end will be called the positive (negative) breakup, at location
$v_{\mathrm{pos}}$ ($v_{\mathrm{neg}}$) in the positive (negative) region,
together with the final breakup forming the positive (negative) hadron.
The first implemented method consists in projecting the positive/negative
hadron four-momentum on to the longitudinal and transverse direction vectors
of the positive/negative region \cite{Sjostrand:1984ic}. If the two old
breakups are in the same region then that is also the region of the final
vertex. This situation is exemplified in Fig.~\ref{fig:finaltwoproj},
where the green point is the final breakup, the red and blue points
are the old breakups, which correspond to the endpoints of the final
region, and the grey squares represent the two final hadrons
created. As can be seen in the figure, the $x^{\pm}$ fraction of the
positive hadron are $x_{p,\mathrm{Had}}^{\pm}$ and the $\hx^{\pm}$ of the
positive breakup are $\hx_{p,\mathrm{Old}}^{\pm}$. Then, the $\hx_f^{\pm}$
of the final breakup can be obtained as
\begin{equation}
\begin{gathered}
\hx_f^+ = \hx_{p,\mathrm{Old}}^+ - x_{p,\mathrm{Had}}^+ ~,\\
\hx_f^- = \hx_{p,\mathrm{Old}}^- + x_{p,\mathrm{Had}}^- ~.
\end{gathered}
\label{eq:finaltwopositive}
\end{equation}
The same procedure can be followed with the negative breakup and the
negative hadron, giving
\begin{equation}
\begin{gathered}
\hx_f^+ = \hx_{n,\mathrm{Old}}^+ + x_{n,\mathrm{Had}}^+ ~,\\
\hx_f^- = \hx_{n,\mathrm{Old}}^- - x_{n,\mathrm{Had}}^- ~.
\end{gathered}
\label{eq:finaltwonegative}
\end{equation}
In the general case, the solutions of eq.~(\ref{eq:finaltwopositive})
and eq.~(\ref{eq:finaltwonegative}) will agree only if the positive and
negative regions coincide. But the projection method can also be used
when the positive and negative regions are different, in which case the
longitudinal momentum of the positive or negative hadron is projected
on the corresponding region, using either
eq.~(\ref{eq:finaltwopositive}) or eq.~(\ref{eq:finaltwonegative}).
If either of these give projected values $0 \leq \hx^{\pm}_f \leq 1$
then a solution has been found in the respective region, and we are done.
If not, the search continues.

\begin{figure}[tp]
\centering
\includegraphics[scale=0.35]{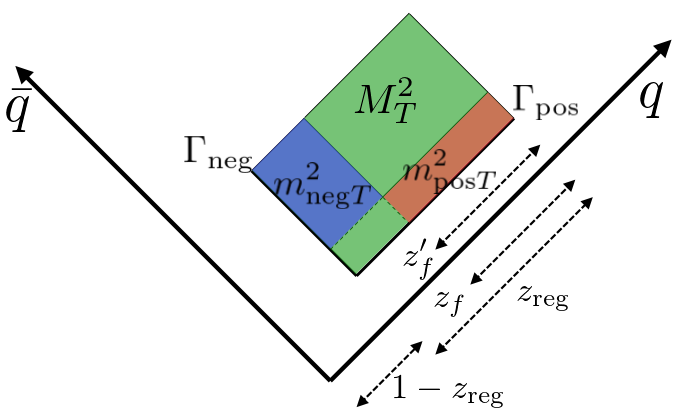}
\caption{Final breakup and final two hadrons of a $\q\qbar$
system. The blue and red areas represent the two final hadrons while
the green area corresponds to the final region.}
\label{fig:finaltwozhad}
\end{figure}

One of the main complications to obtaining the $\hx^{\pm}_f$ values is
that the $z$ value of the final breakup is not calculated, since it is
not needed in the energy-momentum picture. If the projection method fails,
the $z$ value can be calculated from the $\Gamma$ of the old breakups and
the transverse mass of the final region, i.e.\ of the final two hadrons
combined. These variables are depicted in Fig.~\ref{fig:finaltwozhad},
along with $z_f$ and $z'_f$, which correspond to the $z^+$ fractions
of the positive hadron in the real region and the $z^+$ fraction
with respect to the final region, i.e, when the $\tdp^+$ and $\tdp^-$
of the final region are normalized to unity. The $z_{\mathrm{reg}}$
variable in the figure represents the $z^+$ fraction taken from the
negative breakup if the final breakup was not created. This value can
be calculated from the relation given by eq.~(\ref{eq:newoldgammas}),
which in this case gives
\begin{equation}
\Gamma_{\mathrm{neg}} =  (1-z_{\mathrm{reg}}) (\Gamma_{\mathrm{pos}}
+ \frac{M_{\perp}^2}{z_{\mathrm{reg}}}) ~,
\end{equation}
where the final breakup was not taken into account. From
this relation, the value of $z_{\mathrm{reg}}$ is found to be
\begin{equation}
z_{\mathrm{reg}} = \frac{\sqrt{ (M_{\perp}^2 + \Gamma_{\mathrm{neg}}
- \Gamma_{\mathrm{pos}})^2 + 4 M_{\perp}^2 \Gamma_{\mathrm{pos}}}
- (M_{\perp}^2 + \Gamma_{\mathrm{neg}} - \Gamma_{\mathrm{pos}}) }
{2\Gamma_{\mathrm{pos}}} ~.
\end{equation}

During the fragmentation process in \textsc{Pythia} only the fractions
$z'_f$ are determined by considering $z_{\mathrm{reg}} = 1$, as stated
previously. Hence, the $z_f = z_f^+$ fractions can be calculated from
$z'_f$ and $z_{\mathrm{reg}}$ using the relation
$z_f = z'_f \ z_{\mathrm{reg}}$. Note that $z_f = z'_f$ if
$z_{\mathrm{reg}} = 1$, as expected. The same process can be followed
to calculate $z_f^-$, swapping the variables
$\Gamma_{\mathrm{pos}}$ and $\Gamma_{\mathrm{neg}}$. Once the
$z_f$ are known, the $\hx^{\pm}$ fractions of the final breakup
can be determined from eq.~(\ref{eq:relationzx}) and its
space--time location deduced as in section~\ref{sec:spacetime}.

\begin{figure}[tp]
\centering
\includegraphics[scale=0.35]{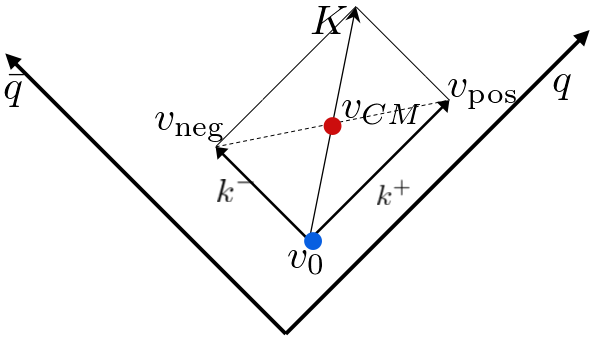}
\caption{Final region of a $\q\qbar$ system. The origin of the
region is represented by $v_0$ while $v_{\mathrm{CM}}$ stands
for the location of the CM of the region. The four-momentum of the
final region is defined as $K$. As in other cases,
$v_{\mathrm{pos}}$ and $v_{\mathrm{neg}}$ represent the old breakups.}
\label{fig:finaltwospecial}
\end{figure}

Although the last procedure succeeds in the large majority of cases,
sometimes the region of the final breakup cannot be found. Then the
location of the final breakup is calculated from the old
breakups by determining a space--time location of the origin of the
artificial final region  used in the energy--momentum picture.
As can be deduced from Fig.~\ref{fig:finaltwospecial}, the
space--time location of the CM of the final region is defined by
$v_{\mathrm{CM}} = (v_{\mathrm{pos}} + v_{\mathrm{neg}}) /2$. This final
region can be treated as a $\q\qbar$ system with four-momentum
$K = k^+ + k^-$, where $k^{\pm}$ are the four-momenta vectors of the
two endpoints. Following the same approach as used to derive the early
hadron production point (see section~\ref{sec:breakuppoints}), the
space--time location of origin of the final region is given by
\begin{equation}
v_0 = v_{\mathrm{CM}} - \frac{1}{2} K.
\end{equation}
Considering $\hx'^{\pm}_f$ to be the $\hx^{\pm}$ fractions of the final
breakup in the final region, the space--time location of the final
breakup can then be calculated as
\begin{equation}
v_f = v_0 + \hx'^+ k^+ + \hx'^- k^- =
v_{\mathrm{CM}} + (\hx'^+ - \frac{1}{2}) k^+ + (\hx'^- - \frac{1}{2}) k^- ~.
\label{eq:specialfinalsame}
\end{equation}
Eq.~(\ref{eq:specialfinalsame}) holds in the transverse rest frame,
when the final region system is not evolving in time. That is not the
case if the two old breakups are in different regions. In order to find
an expression valid for all the cases, we define the variable
$r = \sqrt{l^2/K^2}$, with $l^2 = - (v_{\mathrm{neg}} - v_{\mathrm{pos}})^2$,
to quantify how much the final string system differs from the system in
the transverse rest frame. Then, the origin of the final region in the
string system is determined by $v_0 = v_{\mathrm{CM}} - Kr/2$ and the
general expression for the space--time location of the final breakup is
\begin{equation}
v_f = v_{\mathrm{CM}} + (\hx'^+ - \frac{r}{2}) k^+
+ (\hx'^- - \frac{r}{2}) k^- ~.
\label{eq:specialfinaldifferent}
\end{equation}

The methods presented in this section are implemented in \textsc{Pythia}
to obtain the space--time location of the final breakup. First the
projection method is executed from the $\q$ or positive side. If it
fails, the same method from the negative side is carried out. Whenever
the projection method fails, the $z^+$ value is calculated. In case of
failure with this, the same method is carried out to calculate $z^-$.
If none of the previous methods work, the space--time location of the
final breakup is determined by eq.~(\ref{eq:specialfinaldifferent}).
The different procedures are carried out in order of decreasing accuracy.

\section{Correction to non-physical situations}
\label{sec:nonphysical}

As mentioned in section~\ref{sec:howpointsobtained}, by definition
$0 < \hx^{\pm} < 1$. Although this should always be the case,
\textsc{Pythia} allows values outside that range whenever a step is
taken from one region to a new one. Since the $\hx^{\pm}$
fractions were only used to determine the energy and momentum of the
hadrons, stepping outside the allowed range was not a problem, as long
as the hadron energy was positive. Nevertheless, fractions outside the
$0 < \hx^{\pm} < 1$ range are a significant problem in the space--time
picture, since they can lead to negative times or negative squared
invariant times of the breakup locations. In order to address these
unphysical situations, one of two corrections is applied in the space--time
implementation, before adding the region offset, the smearing in
transverse space and the massive correction.

The first option consists in adjusting the old space--time location of
the breakup by a fraction of the region four-momentum, such that the
new squared invariant time is equal to zero. Then, the expression of
the new breakup location is determined by
\begin{equation}
v_{\mathrm{new}} = v_{\mathrm{old}} + \xi p_{\mathrm{reg}} ~,
\end{equation}
where $\xi$ is found by requiring $v_{\mathrm{new}}^2 = 0$.

In the second option, $\hx^{\pm}$ fractions outside the $0 < \hx^{\pm} < 1$
are set to the value at the closest border, i.e.\ 0 or 1, and the
space--time location is recalculated according to
eq.~(\ref{eq:spacetimebreakup}).

The option adopted is the one that gives the smallest change of the
space--time breakup location.

\end{document}

%% file: setup.tex
\usepackage[utf8]{inputenc}
\usepackage{graphicx}
\usepackage{caption}
\usepackage{subfig}
\usepackage{amsmath}
\usepackage{amssymb}
\usepackage{multirow}
\usepackage{color}

\renewcommand{\b}{{\mathrm b}}
\renewcommand{\c}{{\mathrm c}}
\renewcommand{\d}{{\mathrm d}}
\newcommand{\e}{{\mathrm e}}

\newcommand{\g}{{\mathrm g}}

\newcommand{\n}{{\mathrm n}}
\newcommand{\p}{{\mathrm p}}
\newcommand{\q}{{\mathrm q}}
\newcommand{\s}{{\mathrm s}}

\renewcommand{\u}{{\mathrm u}}

\newcommand{\K}{{\mathrm K}}

\newcommand{\W}{{\mathrm W}}
\newcommand{\Z}{{\mathrm Z}}

\newcommand{\cbar}{\overline{\mathrm c}}
\newcommand{\dbar}{\overline{\mathrm d}}
\newcommand{\nbar}{\overline{\mathrm n}}
\newcommand{\pbar}{\overline{\mathrm p}}
\newcommand{\qbar}{\overline{\mathrm q}}

\newcommand{\sbar}{\overline{\mathrm s}}

\newcommand{\ubar}{\overline{\mathrm u}}

\newcommand{\as}{\alpha_{\mathrm{s}}}

\newcommand{\ECM}{E_{\mathrm{cm}}}

\newcommand{\bp}{\mathbf{p}}

\newcommand{\tdp}{\tilde{p}}
\newcommand{\hx}{\hat{x}}

\newcommand{\mT}{m_{\perp}}

\newcommand{\pT}{p_{\perp}}

\newcommand{\pTo}{p_{\perp 0}}

{\end{list}}
\newcounter{enumct}

\newlength{\abstwidth}


%% file: lutp1818.bbl
\begin{thebibliography}{99}

\bibitem{Campbell:2017hsr}
  J.~Campbell, J.~Huston and F.~Krauss,
  ``The Black Book of Quantum Chromodynamics : A Primer for the LHC Era,''
  Oxford University Press (2017), ISBN: 9780199652747

\bibitem{Sjostrand:1987su}
  T.~Sj\"ostrand and M.~van Zijl,
  Phys.\ Rev.\ D {\bf 36} (1987) 2019.
  doi:10.1103/PhysRevD.36.2019

\bibitem{Sjostrand:2017cdm}
  T.~Sj\"ostrand,
  Adv.Ser.Direct.High Energy Phys. 29 (2018) 191 
  [arXiv:1706.02166 [hep-ph]].

\bibitem{Schael:2013ita}
  S.~Schael {\it et al.} [ALEPH and DELPHI and L3 and OPAL and LEP Electroweak Collaborations],
  Phys.\ Rept.\  {\bf 532} (2013) 119
  doi:10.1016/j.physrep.2013.07.004
  [arXiv:1302.3415 [hep-ex]].

\bibitem{Buckley:2011ms}
  A.~Buckley {\it et al.},
  Phys.\ Rept.\  {\bf 504} (2011) 145
  doi:10.1016/j.physrep.2011.03.005
  [arXiv:1101.2599 [hep-ph]].

\bibitem{Sjostrand:2006za}
  T.~Sj\"ostrand, S.~Mrenna and P.~Z.~Skands,
  JHEP {\bf 0605} (2006) 026
  doi:10.1088/1126-6708/2006/05/026
  [hep-ph/0603175].

\bibitem{Sjostrand:2014zea}
  T.~Sj\"ostrand, S.~Ask, J.~R.~Christiansen, R.~Corke, N.~Desai, P.~Ilten,
  S.~Mrenna and S.~Prestel {\it et al.},
  Comput.\ Phys.\ Commun.\  {\bf 191} (2015) 159
  [arXiv:1410.3012 [hep-ph]].

\bibitem{Bahr:2008pv}
  M.~B\"ahr {\it et al.},
  Eur.\ Phys.\ J.\ C {\bf 58} (2008) 639
  doi:10.1140/epjc/s10052-008-0798-9
  [arXiv:0803.0883 [hep-ph]].

\bibitem{Bellm:2015jjp}
  J.~Bellm {\it et al.},
  Eur.\ Phys.\ J.\ C {\bf 76} (2016) no.4,  196
  doi:10.1140/epjc/s10052-016-4018-8
  [arXiv:1512.01178 [hep-ph]].

\bibitem{Gleisberg:2008ta}
  T.~Gleisberg, S.~H\"oche, F.~Krauss, M.~Sch\"onherr, S.~Schumann, F.~Siegert and J.~Winter,
  JHEP {\bf 0902} (2009) 007
  doi:10.1088/1126-6708/2009/02/007
  [arXiv:0811.4622 [hep-ph]].

\bibitem{Andersson:1983ia}
  B.~Andersson, G.~Gustafson, G.~Ingelman and T.~Sj\"ostrand,
  Phys.\ Rept.\  {\bf 97} (1983) 31.

\bibitem{Webber:1983if}
  B.~R.~Webber,
  Nucl.\ Phys.\ B {\bf 238} (1984) 492.
  doi:10.1016/0550-3213(84)90333-X

\bibitem{ALICE:2017jyt}
  J.~Adam {\it et al.} [ALICE Collaboration],
  Nature Phys.\  {\bf 13} (2017) 535
  doi:10.1038/nphys4111
  [arXiv:1606.07424 [nucl-ex]].

\bibitem{Khachatryan:2010gv}
  V.~Khachatryan {\it et al.} [CMS Collaboration],
  JHEP {\bf 1009} (2010) 091
  doi:10.1007/JHEP09(2010)091
  [arXiv:1009.4122 [hep-ex]].

\bibitem{Khachatryan:2015lva}
  V.~Khachatryan {\it et al.} [CMS Collaboration],
  Phys.\ Rev.\ Lett.\  {\bf 116} (2016) no.17,  172302
  doi:10.1103/PhysRevLett.116.172302
  [arXiv:1510.03068 [nucl-ex]].

\bibitem{Aad:2015gqa}
  G.~Aad {\it et al.} [ATLAS Collaboration],
  Phys.\ Rev.\ Lett.\  {\bf 116} (2016) no.17,  172301
  doi:10.1103/PhysRevLett.116.172301
  [arXiv:1509.04776 [hep-ex]].

\bibitem{Ortiz:2013yxa}
  A.~Ortiz Velasquez, P.~Christiansen, E.~Cuautle Flores, I.~Maldonado Cervantes and G.~Paić,
  Phys.\ Rev.\ Lett.\  {\bf 111} (2013) no.4,  042001
  doi:10.1103/PhysRevLett.111.042001
  [arXiv:1303.6326 [hep-ph]].

\bibitem{Abelev:2014qqa}
  B.~B.~Abelev {\it et al.} [ALICE Collaboration],
  Eur.\ Phys.\ J.\ C {\bf 75} (2015) no.1,  1
  doi:10.1140/epjc/s10052-014-3191-x
  [arXiv:1406.3206 [nucl-ex]].

\bibitem{Khachatryan:2016txc}
  V.~Khachatryan {\it et al.} [CMS Collaboration],
  Phys.\ Lett.\ B {\bf 765} (2017) 193
  doi:10.1016/j.physletb.2016.12.009
  [arXiv:1606.06198 [nucl-ex]].

\bibitem{BraunMunzinger:2007zz}
  P.~Braun-Munzinger and J.~Stachel,
  Nature {\bf 448} (2007) 302.
  doi:10.1038/nature06080

\bibitem{Busza:2018rrf}
  W.~Busza, K.~Rajagopal and W.~van der Schee,
  Ann.Rev.Nucl.Part.Sci. 68 (2018) 339
  [arXiv:1802.04801 [hep-ph]].

\bibitem{Nagle:2018nvi}
  J.~L.~Nagle and W.~A.~Zajc,
  Ann.Rev.Nucl.Part.Sci. 68 (2018) 211
  [arXiv:1801.03477 [nucl-ex]].

\bibitem{Werner:2014xoa}
  K.~Werner, B.~Guiot, I.~Karpenko and T.~Pierog,
  Nucl.\ Phys.\ A {\bf 931} (2014) 83
  doi:10.1016/j.nuclphysa.2014.08.093
  [arXiv:1411.1048 [nucl-th]].

\bibitem{Bierlich:2014xba}
  C.~Bierlich, G.~Gustafson, L.~L\"onnblad and A.~Tarasov,
  JHEP {\bf 1503} (2015) 148
  doi:10.1007/JHEP03(2015)148
  [arXiv:1412.6259 [hep-ph]].

\bibitem{Bierlich:2016vgw}
  C.~Bierlich, G.~Gustafson and L.~L\"onnblad,
  arXiv:1612.05132 [hep-ph].

\bibitem{Fischer:2016zzs}
  N.~Fischer and T.~Sj\"ostrand,
  JHEP {\bf 1701} (2017) 140
  doi:10.1007/JHEP01(2017)140
  [arXiv:1610.09818 [hep-ph]].

\bibitem{Sjostrand:1984ic}
  T.~Sj\"ostrand,
  Nucl.\ Phys.\ B {\bf 248} (1984) 469.
  doi:10.1016/0550-3213(84)90607-2

\bibitem{Gallmeister:2005ad}
  K.~Gallmeister and T.~Falter,
  Phys.\ Lett.\ B {\bf 630} (2005) 40
  doi:10.1016/j.physletb.2005.08.135
  [nucl-th/0502015].

\bibitem{Artru:1974hr}
  X.~Artru and G.~Mennessier,
  Nucl.\ Phys.\ B {\bf 70} (1974) 93.
  doi:10.1016/0550-3213(74)90360-5

\bibitem{Andersson:1983jt}
  B.~Andersson, G.~Gustafson and B.~S\"oderberg,
  Z.\ Phys.\ C {\bf 20} (1983) 317.
  doi:10.1007/BF01407824

\bibitem{Andersson:1984af}
  B.~Andersson, G.~Gustafson and T.~Sj\"ostrand,
  Phys.\ Scripta {\bf 32} (1985) 574.
  doi:10.1088/0031-8949/32/6/003

\bibitem{Eden:1996xi}
  P.~Ed\'en and G.~Gustafson,
  Z.\ Phys.\ C {\bf 75} (1997) 41
  doi:10.1007/s002880050445
  [hep-ph/9606454].

\bibitem{tHooft:1973alw}
  G.~'t Hooft,
  Nucl.\ Phys.\ B {\bf 72} (1974) 461.
  doi:10.1016/0550-3213(74)90154-0

\bibitem{Sjostrand:2004pf}
  T.~Sj\"ostrand and P.~Z.~Skands,
  JHEP {\bf 0403} (2004) 053
  doi:10.1088/1126-6708/2004/03/053
  [hep-ph/0402078].

\bibitem{Patrignani:2016xqp}
  C.~Patrignani {\it et al.} [Particle Data Group],
  Chin.\ Phys.\ C {\bf 40} (2016) no.10,  100001.
  doi:10.1088/1674-1137/40/10/100001

\bibitem{Sjostrand:2002ip}
  T.~Sj\"ostrand and P.~Z.~Skands,
  Nucl.\ Phys.\ B {\bf 659} (2003) 243
  doi:10.1016/S0550-3213(03)00193-7
  [hep-ph/0212264].

\bibitem{Christiansen:2015yqa}
  J.~R.~Christiansen and P.~Z.~Skands,
  JHEP {\bf 1508} (2015) 003
  doi:10.1007/JHEP08(2015)003
  [arXiv:1505.01681 [hep-ph]].

\bibitem{Dokshitzer:1991wu}
  Y.~L.~Dokshitzer, V.~A.~Khoze, A.~H.~Mueller and S.~I.~Troian,
  ``Basics of perturbative QCD,''
  Gif-sur-Yvette, France: Ed. Frontieres (1991) 274 p. (Basics of)

\bibitem{Corke:2009tk}
  R.~Corke and T.~Sjostrand,
  JHEP {\bf 1001} (2010) 035
  doi:10.1007/JHEP01(2010)035
  [arXiv:0911.1909 [hep-ph]].

\bibitem{Bialas:1986cf}
  A.~Bialas and M.~Gyulassy,
  Nucl.\ Phys.\ B {\bf 291} (1987) 793.
  doi:10.1016/0550-3213(87)90496-2

\bibitem{Skands:2014pea}
  P.~Skands, S.~Carrazza and J.~Rojo,
  Eur.\ Phys.\ J.\ C {\bf 74} (2014) no.8,  3024
  doi:10.1140/epjc/s10052-014-3024-y
  [arXiv:1404.5630 [hep-ph]].

\bibitem{Zhang:2017esm}
  Y.~X.~Zhang {\it et al.},
  Phys.\ Rev.\ C {\bf 97} (2018) no.3,  034625
  doi:10.1103/PhysRevC.97.034625
  [arXiv:1711.05950 [nucl-th]].

\bibitem{Bass:1998ca}
  S.~A.~Bass {\it et al.},
  Prog.\ Part.\ Nucl.\ Phys.\  {\bf 41} (1998) 255
  doi:10.1016/S0146-6410(98)00058-1
  [nucl-th/9803035].

\bibitem{Weil:2016zrk}
  J.~Weil {\it et al.},
  Phys.\ Rev.\ C {\bf 94} (2016) no.5,  054905
  doi:10.1103/PhysRevC.94.054905
  [arXiv:1606.06642 [nucl-th]].

\end{thebibliography}
